\documentclass[aps,prb,reprint,superscriptaddress,amsmath,amssymb,floatfix]{revtex4-2}

\usepackage{mathtools}
\usepackage{bm}
\usepackage{times}
\usepackage{graphicx}
\graphicspath{{Figs/}}
\usepackage{xcolor}
\definecolor{revisionteal}{RGB}{0,105,105}
\definecolor{revisionbrown}{RGB}{139,69,19}
\usepackage{amsthm}
\usepackage{mathrsfs}
\usepackage[colorlinks,linkcolor=blue,citecolor=blue,urlcolor=blue]{hyperref}

\begin{document}

\title{Amplified Memory and Finite-Time Regularity in Driven Non-Hermitian Systems}

\author{H. Yavartanoo}
\affiliation{Beijing Institute of Mathematical Sciences and Applications (BIMSA),
Huairou District, Beijing 101408, China}

\author{R. Jafari}
\affiliation{Department of Physics, Institute for Advanced Studies in Basic Sciences (IASBS),
Zanjan 45137-66731, Iran}
\affiliation{School of Quantum Physics and Matter Science,
Institute for Research in Fundamental Sciences (IPM),
19395-5531 Tehran, Iran}

\author{Alireza Akbari}
\affiliation{Beijing Institute of Mathematical Sciences and Applications (BIMSA),
Huairou District, Beijing 101408, China}
\affiliation{Max Planck Institute for the Chemical Physics of Solids,
D-01187 Dresden, Germany}

\date{\today}
\begin{abstract}
We study the roles of broken spectra and exceptional points in finite-time driven non-Hermitian fermionic dynamics. We compute the Nambu biorthogonal correlation matrix for this purpose. The imbalanced-pairing Kitaev chain serves as our concrete realization.
Transient passage through a broken-spectrum region amplifies preparation memory. The effect survives when the final Hamiltonian returns to a real-spectrum regime. Negative imbalance forces spectral nonpositivity in the static endpoint state. We isolate the genuine drive history by subtracting out this baseline, leaving an excess that is strictly controlled by the accumulated imaginary-energy action and persists over the entire post-ramp time window. A connected longitudinal correlation mirrors this physics. Its slow-ramp growth tracks the corresponding doubled action. Unstable sectors instead continue amplifying post-ramp if the drive halts inside the broken-spectrum region.
Exceptional points yield distinct physics. The finite-time propagator and subsystem correlation matrix remain entirely regular near an exceptional endpoint, even as the quasiparticle gap exhibits its characteristic square-root closing.
This finite-time regularity reflects the analyticity of the matrix evolution in the endpoint parameter; a diagonalizable endpoint is strictly not required.
A diverging long-time crossover eventually reveals the exceptional scale. We halt the drive exactly at the exceptional point to find that the correlation projector and the connected longitudinal correlation share an identical ballistic front. The subsystem saturation length establishes a distinct but comparable spatial scale. Memory and exceptional-endpoint scalings show no divergence across the tested negative-imbalance range. Memory scaling is fixed by the drive and remains insensitive to the specific choice of real-spectrum final endpoint.
\end{abstract}

\maketitle

\section{Introduction}
\label{sec:introduction}

Entanglement diagnoses quantum many-body correlations across both equilibrium and driven regimes. For quadratic Hermitian fermions, two-point Gaussian correlators fix the reduced density matrix entirely; one extracts the entanglement spectrum and entropy directly without handling the full state vector~\cite{Peschel2003,PeschelEisler2009}. Identical constructions apply to topological superconductors~\cite{Fidkowski2010}.
It would be intriguing to apply this correlation-based methodology to non-Hermitian systems.
Non-Hermitian systems require a biorthogonal framework because right and left eigenstates span different bases~\cite{Brody2014}. This non-Hermiticity drives complex band spectra, exceptional points, and non-Hermitian topology~\cite{Kawabata2019,Ashida2020,Bergholtz2021}. Different definitions of the reduced state isolate different physical information~\cite{Herviou2019,Chang2020,Guo2021,Tu2022}.
Non-Hermitian systems require a biorthogonal framework because right and left eigenstates span different bases~\cite{Brody2014}. This non-Hermiticity drives complex band spectra, exceptional points, and non-Hermitian topology~\cite{Kawabata2019,Ashida2020,Bergholtz2021}. Different definitions of the reduced state isolate different physical information~\cite{Herviou2019,Chang2020,Guo2021,Tu2022}.
Static biorthogonal entanglement in non-interacting systems is well understood. Prior studies link the single-particle entanglement spectrum to non-Hermitian topology; the resulting entropy can become negative or complex and host non-unitary critical exponents~\cite{Herviou2019,Chang2020,Guo2021,Tu2022,XueLee2026}. Other works use entanglement to probe dualities, many-body spectra, and quantum phase boundaries~\cite{SciPostPhys.11.1.003,PhysRevLett.130.010401,26jz-zmsv,Agarwal2026}. Disordered systems exhibit similar behavior~\cite{Tozar2026}, as reviewed in Ref.~\cite{Chen2024Review}.

Driven non-Hermitian settings yield richer dynamics.
Exceptional quenches trigger anomalous correlation spreading~\cite{BacsiDora2021}. Entanglement spectra track dynamical topology~\cite{Sayyad2021,Starchl2022}. 
Dynamical observables can also provide signatures of exceptional points~\cite{PhysRevA.110.012226}.
Non-Hermitian spin chains display distinct velocity regimes~\cite{TurkeshiSchiro2023}. The skin effect induces entanglement transitions~\cite{KawabataNumasawaRyu2023}, and dissipation produces similar transitions in Kitaev chains~\cite{Zhou2024}.
Beyond non-Hermitian Hamiltonians, time-dependent studies deploy quantum information metrics (quantum Fisher information, coherence, and magic) to characterize spin and $p$-wave platforms~\cite{PhysRevA.82.052317,Jafari2015,PhysRevA.98.052338,PhysRevA.101.062105,JafariAkbari2021,Ansari2024}. Quenches near exceptional points in non-Hermitian SSH chains produce supersonic modes and multiple light cones~\cite{BacsiDora2021}. Quenches directly into parity-time ($\mathcal{PT}$)-broken phases drive exponential growth in biorthogonal correlators~\cite{LuChang2026}. 
The difference between separately normalized right- and left-state entanglement entropies following a quench can provide a dynamical signature of exceptional points~\cite{Lakkaraju2026}.
Few-body systems and experimental platforms also host defect-assisted entanglement generation~\cite{Li2023Entanglement,Han2023Exceptional}.

Finite-rate sweeps in Hermitian systems follow standard Kibble--Zurek and adiabatic scaling arguments~\cite{Dziarmaga2010}. Recent work indicates that defect production can decouple from equilibrium criticality~\cite{JafariAkbariKZM2026}. Non-Hermitian ramps yield modified Kibble--Zurek exponents near exceptional boundaries~\cite{DoraHeylMoessner2019}. Complex eigenenergies selectively amplify or damp individual modes during slow driving~\cite{WangLangChong2018}. 
Finite-rate biorthogonal dynamics, particularly across transient broken-spectrum regions, remains largely unexplored.

For the imbalanced-pairing Kitaev chain, recent studies track defect generation, defect freezing, and dynamical quantum phase transitions near exceptional boundaries~\cite{JafariKZM2026,Jafari2026}. Two questions remain open. What residual memory of a temporary broken-spectrum passage survives after a finite-rate ramp? When does the singular scale of an exceptional endpoint become visible in dynamics?
We address both questions using the Nambu biorthogonal correlation matrix in a driven imbalanced-pairing Kitaev chain~\cite{Kitaev2001,Li2018}. A transient excursion across a broken-spectrum interval acts as an amplifier. The integrated imaginary splitting accumulated along the ramp sets a finite-rate excess that survives after the Hamiltonian returns to a real-spectrum phase. Connected longitudinal correlations exhibit the identical drive-history dependence; this confirms the memory is not an artifact of the entanglement diagnostic.

Exceptional endpoints behave differently.
The square-root closing of the instantaneous quasiparticle gap does not induce a finite-time divergence in the reduced-state dynamics. The finite-time propagator depends analytically on the endpoint parameter; Hamiltonian defectiveness causes no immediate non-analyticity over compact durations. The exceptional scale appears only as an asymptotic long-time crossover. At an exact exceptional halt, the correlation projector and connected longitudinal correlations expand with the same ballistic front. The subsystem saturation length sets a distinct expanding spatial scale. A drive ending inside the broken phase provides an unstable, amplifying counterpoint. Positive imbalance provides an exact Hermitian benchmark via similarity transformation. These results separate finite-rate preparation memory from exceptional-point regularity and spatial correlation spreading.

\section{Framework and driven model}
\label{sec:general_formalism}

\subsection{Biorthogonal Gaussian correlation matrix}
\label{sec:general_formalism:secA}

  Consider a translationally invariant quadratic fermionic lattice carrying $M$ physical fermionic modes per momentum sector. Up to an overall constant shift, the Nambu representation of the many-body Hamiltonian reads
\begin{equation}
\hat H(t) =\frac{1}{2} \sum_{\bm k\in{\rm BZ}} \hat{\bm\Psi}_{\bm k}^{\dagger} \,\mathcal H_{\bm k}(t)\, \hat{\bm\Psi}_{\bm k},
\label{eq:general_BdG_H}
\end{equation}
where ${\mathcal H}_{\bm k}(t)$ denotes the $2M\times2M$ Bogoliubov--de Gennes (BdG) matrix. The field operators form the standard Nambu spinor
\begin{equation}
\hat{\bm\Psi}_{\bm k} = \left( \hat c_{\bm k,1}, \ldots, \hat c_{\bm k,M}, \hat c_{-\bm k,1}^{\dagger}, \ldots, \hat c_{-\bm k,M}^{\dagger} \right)^T ,
\label{eq:general_nambu_spinor}
\end{equation}
with $\mathcal H_{\bm k}(t)\neq\mathcal H_{\bm k}^{\dagger}(t)$ when Hermiticity is broken. Pure Gaussian states correspond to rank-$M$ correlation projectors in this $2M$-dimensional space. We parameterize this projector by tracking $M$ right vectors $\vert{}R_{a\bm k}(t)\rangle$ alongside the dual left set $\langle L_{a\bm k}(t)\vert{}$ for $a=1,\ldots,M$. Collecting these amplitudes into rectangular blocks yields
\begin{equation}
\begin{aligned}
R_{\bm k}(t) &= \left( |R_{1\bm k}(t)\rangle,\ldots, |R_{M\bm k}(t)\rangle \right), \\ L_{\bm k}^{\dagger}(t) &= \left( \langle L_{1\bm k}(t)|,\ldots, \langle L_{M\bm k}(t)| \right)^{T}.
\end{aligned}
\label{eq:general_LR_frames}
\end{equation}
The matrix dimensions are explicit: $R_{\bm k}$ is $2M\times M$, while $L_{\bm k}^{\dagger}$ is $M\times2M$. These columns and rows contain single-particle Nambu coefficients. They do not represent many-body states. They simply parameterize the time-dependent Bogoliubov amplitudes of the Gaussian state; generally, they deviate from the instantaneous eigenstates of $\mathcal H_{\bm k}(t)$.

  Left and right frames obey the equations of motion
\begin{equation}
\begin{aligned}
i\partial_t R_{\bm k}(t) = \mathcal H_{\bm k}(t)R_{\bm k}(t), \quad -i\partial_t L_{\bm k}^{\dagger}(t) = L_{\bm k}^{\dagger}(t)\mathcal H_{\bm k}(t).
\end{aligned} \label{eq:general_LR_evolution}
\end{equation}
Setting the initial frames biorthonormal ensures this condition holds for all $t$:
\begin{equation}
L_{\bm k}^{\dagger}(t)R_{\bm k}(t)=I_M .
\label{eq:general_bio_norm_time}
\end{equation}
The corresponding momentum-resolved biorthogonal projector is
\begin{equation}
P_{\bm k}(t) = R_{\bm k}(t)L_{\bm k}^{\dagger}(t) = \sum_{a=1}^{M} |R_{a\bm k}(t)\rangle \langle L_{a\bm k}(t)|.
\label{eq:general_Pk}
\end{equation}
The two contractions serve distinct algebraic purposes. While $L_{\bm k}^{\dagger}R_{\bm k}=I_M$ fixes the $M\times M$ identity overlap, $P_{\bm k}(t)$ acts as a $2M\times2M$ projector onto the subspace spanned by the right vectors. The projector identity follows directly:
\begin{equation}
P_{\bm k}^{\,2}(t) = R_{\bm k} \left(L_{\bm k}^{\dagger}R_{\bm k}\right) L_{\bm k}^{\dagger} = P_{\bm k}(t),
\label{eq:general_Pk_projector}
\end{equation}
The spectrum of $P_{\bm k}(t)$ therefore consists strictly of $M$ unit eigenvalues and $M$ zeros.

  Idempotence and biorthonormality do not suffice on their own to enforce genuine fermionic statistics. Under the basis choice of Eq.~(\ref{eq:general_nambu_spinor}), the canonical anticommutation relations impose an algebraic constraint between opposite momenta:
\begin{equation}
P_{\bm k}(t) = I_{2M} - \Xi P_{-\bm k}^{T}(t)\Xi, \qquad \Xi= \begin{pmatrix} 0&I_M\\ I_M&0 \end{pmatrix}.
\label{eq:fermionic_compatibility}
\end{equation}
The physical initial state must satisfy this relation. Because the BdG kernel satisfies the particle-hole symmetry $\mathcal H_{\bm k}=-\Xi\mathcal H_{-\bm k}^{T}\Xi$ in this convention, the dynamics preserves Eq.~(\ref{eq:fermionic_compatibility}) automatically.

Let $\mathcal U_{\bm k}(t)$ denote the single-particle propagator,
\begin{equation}
i\partial_t\mathcal U_{\bm k}(t) = \mathcal H_{\bm k}(t)\mathcal U_{\bm k}(t), \qquad \mathcal U_{\bm k}(0)=I_{2M}.
\label{eq:general_propagator}
\end{equation}
Then
\begin{equation}
\begin{aligned}
R_{\bm k}(t) = \mathcal U_{\bm k}(t)R_{\bm k}(0), \qquad L_{\bm k}^{\dagger}(t) = L_{\bm k}^{\dagger}(0)\mathcal U_{\bm k}^{-1}(t),
\end{aligned} \label{eq:general_frames_propagator}
\end{equation}
and therefore
\begin{equation}
P_{\bm k}(t) = \mathcal U_{\bm k}(t) P_{\bm k}(0) \mathcal U_{\bm k}^{-1}(t).
\label{eq:general_similarity_evolution}
\end{equation}
The projector $P_{\bm k}(t)$ generally drifts away from $P_{\bm k}(0)$, but its rank-$M$ spectral projection property is invariant. In Hermitian problems, one has $\mathcal U_{\bm k}^{-1}=\mathcal U_{\bm k}^{\dagger}$, reducing Eq.~(\ref{eq:general_similarity_evolution}) to unitary conjugation. For non-Hermitian generators, $\mathcal U_{\bm k}$ is not unitary, and the correlation projector undergoes a nonunitary similarity transformation.

\subsection{Subsystem correlation matrix}

We denote the full many-body Gaussian states by $\vert{}\Psi_R(t)\rangle$ and $\langle\Psi_L(t)\vert{}$. These are distinct objects from the single-particle frames $R_{\bm k}(t)$ and $L_{\bm k}^{\dagger}(t)$ defined above.
  The Nambu field $\hat{\bm\Psi}_{\bm k}$ in Eq.~(\ref{eq:general_nambu_spinor}) represents a vector of $2M$ fermionic operators. The dyadic product $\hat{\bm\Psi}_{\bm k}\hat{\bm\Psi}_{\bm k}^{\dagger}$ is an operator-valued $2M\times2M$ matrix with entries given by $\hat\Psi_{\bm k,\alpha}\hat\Psi_{\bm k,\beta}^{\dagger}$. Evaluated in the Gaussian state, the entries of the projector $P_{\bm k}(t)$ match the normalized two-point contractions:
\begin{equation}
\left[P_{\bm k}(t)\right]_{\alpha\beta} = \frac{ \langle\Psi_L(t)| \hat\Psi_{\bm k,\alpha} \hat\Psi_{\bm k,\beta}^{\dagger} |\Psi_R(t)\rangle }{ \langle\Psi_L(t)|\Psi_R(t)\rangle }.
\label{eq:Pk_manybody_relation}
\end{equation}
Equation~(\ref{eq:general_Pk}) builds $P_{\bm k}$ from single-particle Nambu frames; Eq.~(\ref{eq:Pk_manybody_relation}) relates these entries directly to many-body expectations.

  We construct the normalized biorthogonal density operator via~\cite{Brody2014,Qiu2019}
\begin{equation}
\begin{aligned}
\rho^{LR}(t) = \frac{ |\Psi_R(t)\rangle\langle\Psi_L(t)| }{ \langle\Psi_L(t)|\Psi_R(t)\rangle }, \qquad \rho_A^{LR}(t) = \operatorname{Tr}_{\bar A}\rho^{LR}(t).
\end{aligned} \label{eq:rho_LR}
\end{equation}
where $\bar A$ is the spatial complement of subsystem $A$, and the partial trace sums out all modes outside $A$. The overlap $\langle\Psi_L(t)\vert{}\Psi_R(t)\rangle$ is assumed finite. For this pure biorthogonal construction, $\rho^{LR}$ is idempotent and carries unit trace.
Equation~(\ref{eq:rho_LR}) fixes the overlap-normalized convention standard in non-Hermitian mechanics~\cite{Brody2014,Qiu2019}. This guarantees $\operatorname{Tr}\rho^{LR}=1$ together with $(\rho^{LR})^2=\rho^{LR}$.

An alternative scheme encountered in non-Hermitian models normalizes states via standard Dirac norms~\cite{Zhang2025SelfNormal}. Constructing the transition density matrix under that prescription yields
\begin{equation}
\begin{aligned}
\widetilde{\rho}^{LR} = \frac{ |\Psi_R\rangle\langle\Psi_L| }{ \sqrt{ \langle\Psi_L|\Psi_L\rangle \langle\Psi_R|\Psi_R\rangle } }.
\end{aligned} \label{eq:dirac_normalized_transition_operator}
\end{equation}
This Dirac-normalized form differs fundamentally from Eq.~(\ref{eq:rho_LR}). Taking its trace and square gives
\begin{equation}
\begin{aligned}
\operatorname{Tr}\widetilde{\rho}^{LR} \!= \chi \equiv\! \frac{ \langle\Psi_L|\Psi_R\rangle }{ \sqrt{ \langle\Psi_L|\Psi_L\rangle \langle\Psi_R|\Psi_R\rangle } }, \qquad \bigl(\widetilde{\rho}^{LR}\bigr)^2 \!= \chi\,\widetilde{\rho}^{LR}.
\end{aligned} \label{eq:dirac_vs_biorthogonal_normalization}
\end{equation}
The matrix $\widetilde{\rho}^{LR}$ is generally neither idempotent nor unit trace.
When $\chi\neq0$, dividing out the trace recovers the biorthogonal form: $\widetilde{\rho}^{LR}/\operatorname{Tr}\widetilde{\rho}^{LR}=\rho^{LR}$. We work with Eq.~(\ref{eq:rho_LR}) because it matches the Gaussian projector framework and preserves standard expectation values. This is distinct from schemes tracking right and left states independently to compare separate entropies~\cite{Lakkaraju2026}. 
In general, however, $(\rho^{LR})^\dagger\neq\rho^{LR}$ and it is not positive semidefinite with respect to the usual inner product. The reduced block $\rho_A^{LR}$ has unit trace but is generally neither Hermitian nor idempotent.
In the Hermitian limit, both definitions reduce to the standard density matrix.

  Fourier transforming $P_{\bm k}(t)$ yields the real-space correlation matrix. We write the normalized expectation value as
\begin{equation}
\langle \hat O\rangle_{LR} \equiv \frac{ \langle\Psi_L(t)|\hat O|\Psi_R(t)\rangle }{ \langle\Psi_L(t)|\Psi_R(t)\rangle } = \operatorname{Tr}\!\left[\rho^{LR}(t)\hat O\right],
\label{eq:LR_expectation}
\end{equation}
such that the real-space matrix entries become
\begin{equation}
\mathcal C_{ab}(t) = \langle \hat\Psi_a\hat\Psi_b^\dagger \rangle_{LR}.
\label{eq:general_corr}
\end{equation}
Here $a$ and $b$ label spatial and Nambu indices. We restrict attention to subsystem $A$. For a subsystem composed of $L_A$ lattice sites with $M$ orbitals each,
\begin{equation}
\ell=M L_A
\label{eq:subsystem_modes}
\end{equation}
gives the count of physical modes in $A$. The subsystem correlation matrix $\mathcal C_A$ has dimension $2\ell\times2\ell$. In our Kitaev chain calculations, $M=1$, yielding $\ell=L_A$.

  For superconducting systems, the matrix decomposes into blocks:
\begin{equation}
\begin{aligned}
\mathcal C_A = \begin{pmatrix} I_\ell-C_A^T & F_A\\ \overline F_A & C_A \end{pmatrix},
\end{aligned} \label{eq:general_CA}
\end{equation}
with
\begin{equation}
\begin{aligned}
(C_A)_{ij} &= \langle \hat c_i^\dagger \hat c_j\rangle_{LR}, & (F_A)_{ij} &= \langle \hat c_i \hat c_j\rangle_{LR}, \\ (\overline F_A)_{ij} &= \langle \hat c_i^\dagger \hat c_j^\dagger\rangle_{LR}.
\end{aligned} \label{eq:general_normal_anomalous}
\end{equation}
Non-Hermiticity breaks the conjugacy relation: $\overline F_A\neq F_A^\dagger$. Here $C_A$ denotes only the normal correlation sector, whereas $\mathcal C_A$ denotes the doubled Nambu matrix.

  Fermionic anticommutation inside the subsystem imposes
\begin{equation}
\mathcal C_A = I_{2\ell} - \Xi_A\mathcal C_A^T\Xi_A, \qquad \Xi_A= \begin{pmatrix} 0&I_\ell\\ I_\ell&0 \end{pmatrix}.
\label{eq:subsystem_fermionic_compatibility}
\end{equation}
This identity enforces antisymmetry: $F_A^T=-F_A$, and $\overline F_A^T=-\overline F_A$. It also pairs the single-particle entanglement spectrum as $\lambda\leftrightarrow1-\lambda$. The similarity relation $I_{2\ell}-\mathcal C_A=\Xi_A\mathcal C_A^T\Xi_A\sim\mathcal C_A^T$ confirms this pairing directly.
  Because $\mathcal C_A$ is non-Hermitian, we solve for distinct right and left eigenvectors:
\begin{equation}
\begin{aligned}
\mathcal C_A |\varphi_\nu^R\rangle = \lambda_\nu |\varphi_\nu^R\rangle, \qquad \langle\varphi_\nu^L| \mathcal C_A = \lambda_\nu \langle\varphi_\nu^L|.
\end{aligned} \label{eq:general_ent_spectrum}
\end{equation}
The eigenvalues $\{\lambda_\nu\}$ form the single-particle biorthogonal entanglement spectrum. For diagonalizable matrices, eigenvectors satisfy the biorthonormal gauge
\begin{equation}
\langle\varphi_\mu^L|\varphi_\nu^R\rangle = \delta_{\mu\nu}.
\end{equation}
In Hermitian systems, one has $0\preceq\mathcal C_A\preceq I_{2\ell}$, confining eigenvalues to the real segment $[0,1]$. Without Hermiticity, eigenvalues can leave this interval or acquire non-zero imaginary parts.

\subsection{Entanglement diagnostics}

  When $\mathcal C_A$ is diagonalizable, Wick factorization for Gaussian states~\cite{PeschelEisler2009,Herviou2019,Guo2021}, reviewed in Appendix~\ref{app:entropy_branches}, reduces the entropy to single-particle eigenvalues:
\begin{equation}
S_A = \frac{1}{2} \sum_{\nu=1}^{2\ell} \left[ -\lambda_\nu\operatorname{Log}\lambda_\nu - (1-\lambda_\nu)\operatorname{Log}(1-\lambda_\nu) \right],
\label{eq:general_entropy_spectrum}
\end{equation}
In operator form, this evaluates to
\begin{equation}
S_A = -\frac{1}{2} \operatorname{Tr}_{A} \left[ \mathcal C_A\operatorname{Log}\mathcal C_A + \left(I_{2\ell}-\mathcal C_A\right) \operatorname{Log}\left(I_{2\ell}-\mathcal C_A\right) \right].
\label{eq:general_entropy}
\end{equation}
where $\operatorname{Tr}_{A}$ traces over the Nambu space of region $A$, and the factor $1/2$ removes Nambu double-counting. The matrix logarithms are defined spectrally to match Eq.~(\ref{eq:general_entropy_spectrum}).

Complex or negative values of $\lambda_\nu$ force a choice of branch for the logarithm. We evaluate the entropy under two schemes. The principal-branch entropy $S_A^{\rm pr}$ computes each logarithm on its principal branch at each time step independently. The continuous entropy $S_A^{\rm cont}$ initializes on the principal branch at $t=0$ and tracks eigenvalues continuously, shifting across Riemann sheets when crossing branch cuts to avoid discontinuous jumps. Appendix~\ref{app:entropy_branches} provides the tracking algorithm.

  For real spectra, eigenvalues satisfying $\lambda\notin[0,1]$ generate negative many-body weights. We call this condition spectral nonpositivity. We measure this deviation using
\begin{equation}
\mathcal N_A = \frac{1}{2} \sum_{\nu=1}^{2\ell} \left( [-\lambda_\nu]_+ + [\lambda_\nu-1]_+ \right), \qquad [x]_+\equiv\max(x,0).
\label{eq:general_nonpositivity}
\end{equation}
This index $\mathcal N_A$ applies only to real spectra; it is distinct from standard negativity. For real spectra with $\lambda\leftrightarrow1-\lambda$ pairing,
\begin{equation}
\operatorname{Im}S_A^{\rm pr} = \pi\,\mathcal N_A .
\label{eq:ImS_nonpositivity}
\end{equation}
Here, the imaginary part of $S_A^{\rm pr}$ provides no information beyond $\mathcal N_A$.

\subsection{Driven imbalanced-pairing Kitaev chain}
\label{sec:kitaev_model}

We test these dynamical signatures using an imbalanced-pairing Kitaev chain~\cite{Kitaev2001,Li2018}:
\begin{equation}
\begin{aligned}
\hat H(t)=& \frac{w}{2}\sum_{j=1}^{N} \left(\hat c_j^\dagger \hat c_{j+1}+{\rm H.c.}\right) -\mu(t)\sum_{j=1}^{N}\hat c_j^\dagger \hat c_j \\ &-\frac{1}{2}\sum_{j=1}^{N} \left( \Delta \hat c_j^\dagger \hat c_{j+1}^\dagger +\gamma\Delta \hat c_{j+1}\hat c_j \right),
\end{aligned} \label{eq:Hreal}
\end{equation}
where $w$ is the hopping, $\Delta>0$ sets the pairing scale, and $\gamma\in\mathbb{R}$ controls the pairing imbalance. At $\gamma=1$, one recovers the standard Hermitian model.

We apply a fixed gauge rotation to the standard Nambu basis: $\widetilde\Psi_k=(\hat c_k,-i\hat c_{-k}^{\dagger})^{T}$. All subsequent expressions adopt this representation. The rotation simplifies the algebra by making the BdG matrix real without altering the physics or the eigenvalue spectrum of $\mathcal C_A$. The BdG block becomes~\cite{Jafari2026,Li2018}
\begin{equation}
\begin{aligned}
H_k(t)= \begin{pmatrix} \mu_k(t) & \Delta_k\\ \gamma\Delta_k & -\mu_k(t) \end{pmatrix},
\end{aligned} \label{eq:Hk}
\end{equation}
with parameters
\begin{equation}
\begin{aligned}
\mu_k(t)=w\cos k-\mu(t), \qquad \Delta_k=\Delta\sin k .
\end{aligned} \label{eq:muk_deltak}
\end{equation}
The instantaneous dispersion is
\begin{equation}
\begin{aligned}
E_{k,\pm}(t)=\pm\varepsilon_k(t), \qquad \varepsilon_k(t)= \sqrt{\mu_k^2(t)+\gamma\Delta_k^2}.
\end{aligned} \label{eq:spectrum}
\end{equation}
The chemical potential sweeps linearly at rate $v>0$:
\begin{equation}
\begin{aligned}
\mu(t)=\mu_i+vt, \qquad 0\le t\le t_f, \qquad t_f=\frac{\mu_f-\mu_i}{v}.
\end{aligned} \label{eq:ramp}
\end{equation}
Once $t$ reaches $t_f$, the chemical potential stays pinned at $\mu_f$, and the system evolves during the hold time $\tau=t-t_f\ge0$.

Except where noted, calculations fix $w=\Delta=1$ and $\mu_i=-2$. The initial ground state remains gapped and exhibits a real spectrum for all $\gamma>-3$, covering all parameter values discussed here. For finite even $N$, we impose antiperiodic boundary conditions, working with momentum pairs $(k,-k)$ parameterized by $k=(2m-1)\pi/N$ ($m=1,\ldots,N/2$). The grid size $N_k$ controls momentum-space resolution and is independent of the chain length $N$.

We initialize the system in the gapped ground state at $\mu_i$. In our Nambu basis, $P_k(0)=\langle\widetilde\Psi_k\widetilde\Psi_k^\dagger\rangle_{LR}$ projects onto the positive-energy BdG branch. During the sweep, $P_k(t)$ follows the propagated state according to Eq.~(\ref{eq:general_similarity_evolution}), which differs from the instantaneous ground state of $H_k(t)$. Appendix~\ref{app:kitaev_projector_operator} provides explicit formulas for $M=1$. Initializing with the negative-energy projector yields identical entanglement entropies because particle-hole exchange leaves $s(z)=s(1-z)$ invariant, but alters other observables. We retain the positive-energy convention throughout.

  Negative imbalance ($\gamma<0$) introduces exceptional boundaries where $\varepsilon_k^2=0$. At a boundary, the two roots for $\cos k$ coalesce, producing a double zero at $k=\pm k_{\rm EP}$. This determines the exceptional boundary and its momentum:
\begin{equation}
\begin{aligned}
\mu_{\rm EP} = \sqrt{w^2-\gamma\Delta^2}, \qquad \cos k_{\rm EP} = \frac{w}{\mu_{\rm EP}}.
\end{aligned} \label{eq:kep}
\end{equation}
For $\gamma<0$, the spectrum is complex within the window $\vert{}\mu\vert{}<\mu_{\rm EP}$ and real outside it. At $(k_{\rm EP},\mu_{\rm EP})$, the block becomes defective:
\begin{equation}
\begin{aligned}
H_{\rm EP}^2=0, \qquad H_{\rm EP}\neq0.
\end{aligned} \label{eq:nilpotent}
\end{equation}
This forms a nontrivial $2\times2$ Jordan block. To approach the boundary from the real-spectrum sector, we parameterize the endpoint as
\begin{equation}
\begin{aligned}
\mu_f=\mu_{\rm EP}+\delta\mu, \qquad \delta\mu>0.
\end{aligned} \label{eq:continuous_endpoint}
\end{equation}

\subsection{Hermitian-equivalent benchmark for $\gamma>0$}

  For positive pairing imbalance ($\gamma>0$), a local, time-independent similarity transformation maps the model directly to a Hermitian Kitaev chain~\cite{Li2018}. We deploy this transformation to benchmark the driven biorthogonal dynamics. Defining
\begin{equation}
\mathcal S_{\gamma} = \begin{pmatrix} \gamma^{-1/4} & 0\\ 0 & \gamma^{1/4} \end{pmatrix},
\label{eq:Sgamma}
\end{equation}
the BdG Hamiltonian satisfies
\begin{equation}
H_k(t) = \mathcal S_{\gamma}\, h_k(t)\, \mathcal S_{\gamma}^{-1},
\label{eq:similarity_H}
\end{equation}
where
\begin{equation}
h_k(t) = \begin{pmatrix} \mu_k(t) & \sqrt{\gamma}\,\Delta_k\\ \sqrt{\gamma}\,\Delta_k & -\mu_k(t) \end{pmatrix}
\label{eq:effective_Hermitian_H}
\end{equation}
corresponds to a Hermitian Kitaev chain with effective pairing parameter $\Delta_{\rm eff}=\sqrt{\gamma}\,\Delta$.

\begin{figure}[t]
\centering
\includegraphics[width=\columnwidth]{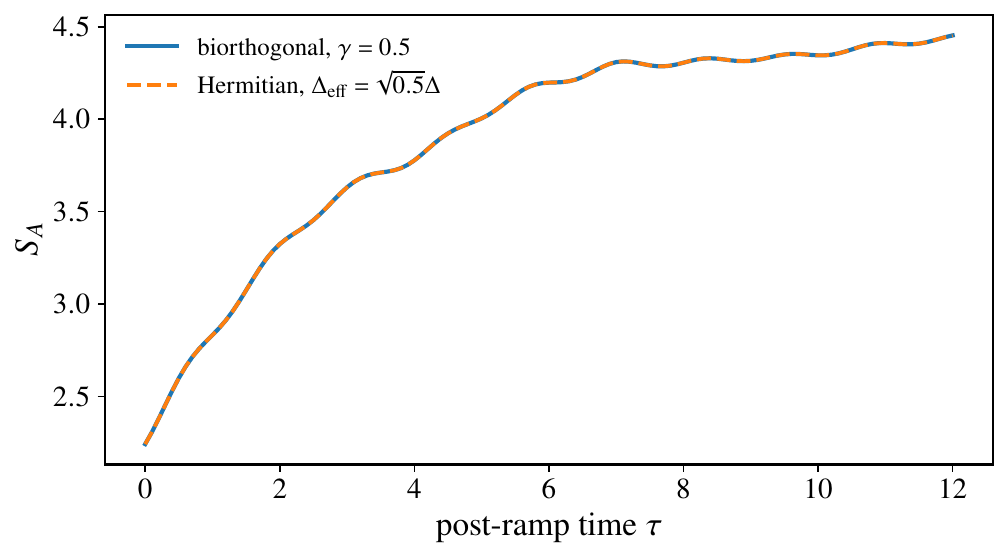}
\caption{
Exact similarity benchmark for $\gamma=0.5$. The biorthogonal entropy of the imbalanced-pairing chain and the entropy of the Hermitian Kitaev chain with $\Delta_{\rm eff}=\sqrt{0.5}\Delta$ coincide throughout the post-ramp evolution. Parameters are {$\mu_i=-2$}, $\mu_f=1.5$, $v=0.5$, and $\ell=10$. The post-ramp oscillations arise from coherent quasiparticle interference in the state prepared by the ramp; the Hermitian benchmark captures them identically.
}
\label{fig:similarity_benchmark}
\end{figure}

  Because $\mathcal S_{\gamma}$ is diagonal and independent of momentum or time, it acts locally in real space. For the mapped initial states, correlation projectors remain related by this similarity transformation at all times. Restricting the state to subsystem $A$ gives
\begin{equation}
\mathcal C_A(t) = \mathcal S_{A,\gamma}\, c_A(t)\, \mathcal S_{A,\gamma}^{-1}, \quad \mathcal S_{A,\gamma} = \begin{pmatrix} \gamma^{-1/4}I_\ell & 0\\ 0 & \gamma^{1/4}I_\ell \end{pmatrix},
\label{eq:similarity_CA}
\end{equation}
where $c_A(t)$ is the subsystem correlation matrix of the Hermitian chain. Since similarity preserves the spectrum, $\mathcal C_A(t)$ and $c_A(t)$ share identical eigenvalues. The resulting entropies match identically:
\begin{equation}
S_A^{\rm bio}(t;\gamma,\Delta) = S_A^{\rm Herm} \left(t;\Delta_{\rm eff}=\sqrt{\gamma}\Delta\right), \qquad \gamma>0.
\label{eq:similarity_entropy}
\end{equation}
For positive imbalance, the biorthogonal dynamics is strictly Hermitian-equivalent. Figure~\ref{fig:similarity_benchmark} demonstrates this correspondence for $\gamma=0.5$.

\section{Amplified finite-rate memory at a real-spectrum endpoint}
\label{sec:memory}

\subsection{Static baseline and spectral nonpositivity}
\label{sec:memory_baseline}

We fix $\gamma<0$ and drive the system across the broken-spectrum region into the unbroken domain beyond the positive exceptional boundary. The final Hamiltonian is diagonalizable and possesses a real spectrum.
The post-ramp projector evolves as
\begin{equation}
\begin{aligned}
P_k(\tau) = e^{-iH_k(\mu_f)\tau} P_k(t_f) e^{+iH_k(\mu_f)\tau}.
\end{aligned} \label{eq:postramp_projector}
\end{equation}
For $\gamma=-0.5$, the positive boundary sits at $\mu_{\rm EP}=\sqrt{1.5}\simeq 1.225$.

Choosing $\mu_f=1.5>\mu_{\rm EP}$ yields real subsystem entanglement eigenvalues outside $[0,1]$ for both the ramp-prepared state and the static reference. The final Hamiltonian spectrum remains entirely real. Spectral reality follows directly from an indefinite pseudo-Hermitian metric obeyed by the Bloch Hamiltonian. For real $\gamma<0$,
\begin{equation}
H_k^\dagger\eta=\eta H_k, \qquad \eta= \begin{pmatrix} \gamma&0\\ 0&1 \end{pmatrix},
\label{eq:eta_pseudo_hermiticity}
\end{equation}
and the positive-energy ground-state projector on the left side ($\mu_i<-\mu_{\rm EP}$) carries a definite metric signature. The dual evolution generated by $\mathcal U_{\bm k}^{-1}(t)$ preserves these signature relations. Restricting the real-space matrix to subsystem $A$ establishes
\begin{equation}
\operatorname{spec}\mathcal C_A(t) \subset (-\infty,0]\cup[1,\infty).
\label{eq:CA_spectral_constraint}
\end{equation}
Appendix~\ref{app:spectral_reality} details the proof. The drive rescales the magnitude of the response while leaving the reality of the single-particle entanglement spectrum intact.

Complementary spectral pairs satisfy $\lambda_\nu<0$ alongside $1-\lambda_\nu>1$. Through the Gaussian mode factorization of Appendix~\ref{app:entropy_branches}, these outlying eigenvalues yield negative many-body weights; the state exhibits spectral nonpositivity. At fixed momentum, evolution is purely oscillatory because post-quench quasiparticle energies are real. In the thermodynamic limit, momentum integration induces dephasing across subsystem observables, precluding persistent oscillations.

Spectral nonpositivity alone cannot diagnose dynamical memory. The finite-rate excitation must be measured relative to the static reference state at the same endpoint. This distinction separates our protocol from sudden quenches terminating directly at exceptional points or inside broken phases~\cite{BacsiDora2021,LuChang2026}. Ref.~\cite{LuChang2026} obtains exponential growth in biorthogonal measures following a quench into a $\mathcal{PT}$-broken phase, linking this divergence to nonpositive density matrices. Here the ramp traverses the complex-energy window transiently and measures the preparation memory retained deep inside the real-spectrum phase.

Our backward-propagated dual frame differs from the final-state dual employed to define positive DQPT rates in Ref.~\cite{Jafari2026}. The resulting spectral weights are not transition probabilities. We neither invoke the $p_k=1/2$ condition nor extract dynamical phase boundaries from this construction.

\subsection{Prepared-state excess and accumulated action}
\label{sec:prep_memory}

 Let $\mathcal N_A^{\rm stat}(\mu_f)$ denote the nonpositivity of the static biorthogonal ground state at $\mu_f$, evaluated using the positive-energy branch. The value reached at the end of the ramp is $\mathcal N_A^{\rm prep}(v)=\mathcal N_A(t_f;v)=\mathcal N_A(\tau=0;v)$. We define the excess nonpositivity and its initial value as
\begin{equation}
\begin{aligned}
\Delta\mathcal N_A(\tau;v) &\equiv \mathcal N_A(\tau;v) - \mathcal N_A^{\rm stat}(\mu_f), \qquad \tau\ge0,
\end{aligned} \label{eq:excess_nonpositivity_time}
\end{equation}
\begin{equation}
\begin{aligned}
\Delta\mathcal N_A^{\rm prep}(v) &\equiv \Delta\mathcal N_A(0;v) = \mathcal N_A^{\rm prep}(v) - \mathcal N_A^{\rm stat}(\mu_f).
\end{aligned} \label{eq:excess_nonpositivity_prep}
\end{equation}
This difference isolates the non-adiabatic contribution generated by driving away from the static baseline. It can take either sign. Evaluating the static reference at $\gamma=-0.5$, $\mu_f=1.5$, and $\ell=10$ gives $\mathcal N_A^{\rm stat}\simeq0.06593$.

Inside the broken-spectrum interval, the instantaneous dispersion is imaginary: $E_{k,\pm}=\pm i\kappa_k$ with $\kappa_k>0$. The off-diagonal components of the correlation projector depend on the energy difference $E_{k,+}-E_{k,-}=2i\kappa_k$. The dominant amplification factor accumulated during the sweep is $\exp[2\int_{\rm broken}dt\,\kappa_k(t)]$, where the integral covers the time window where momentum $k$ develops an imaginary splitting. For a linear sweep with $dt=d\mu/v$, we define the single-particle action
\begin{equation}
\mathcal A(k) = 2\int_{\rm broken}d\mu\,\kappa_k(\mu),
\label{eq:kitaev_ramp_action_general}
\end{equation}
yielding the exponential growth factor $\exp[\mathcal A(k)/v]$~\cite{WangLangChong2018}.
 For linear driving, the imaginary energy component reads
\begin{equation}
\kappa_k(\mu) = \sqrt{ |\gamma|\Delta^2\sin^2k - (\mu-w\cos k)^2 }.
\label{eq:kappa_ramp}
\end{equation}
When the ramp traverses the full imaginary interval for a given mode, Eq.~(\ref{eq:kitaev_ramp_action_general}) integrates to
\begin{equation}
\frac{\mathcal A(k)}{v} = \frac{2}{v} \int_{\rm broken}d\mu\,\kappa_k(\mu) = \frac{\pi|\gamma|\Delta^2\sin^2k}{v}.
\label{eq:ramp_action}
\end{equation}
This expression follows from the semicircle integral over $x=\mu-w\cos k$ within $x^2\le {\vert{}\gamma\vert{}}\Delta^2 \sin^2 k$. The action is maximized at the two symmetry-related momenta $k_0=\pm\pi/2$, where $\mathcal A_{\max}=\pi\vert{}\gamma\vert{}\Delta^2$. Expanding about the saddle, $\mathcal A(k)=\mathcal A_{\max}-\pi\vert{}\gamma\vert{}\Delta^2(k-k_0)^2+\cdots$, reveals that momentum integration introduces an asymptotic prefactor of $\sqrt v$.
The static term is independent of $v$. Subtracting it leaves the exponential action unchanged. Provided the saddle contribution survives the spatial projection onto $A$, the prepared excess scales as
\begin{equation}
\Delta\mathcal N_A^{\rm prep}(v) \sim A_0\sqrt v\, \exp\left[ \frac{\pi|\gamma|\Delta^2}{v} \right],
\label{eq:velocity_scaling}
\end{equation}
where $A_0$ is a velocity-independent prefactor.

The asymptotic derivation is given in Appendix~\ref{app:slow_ramp_asymptotics}. The momentum envelope integrates to a modified Bessel function, confirming the $\sqrt v$ dependence and separating it from the geometric subsystem factor. Compensating the slow-ramp numerical data by $\sqrt v$ yields a constant plateau, showing that the saddle governs the subsystem eigenvalues.

\begin{figure}[t]
\centering
\includegraphics[width=0.5\textwidth]{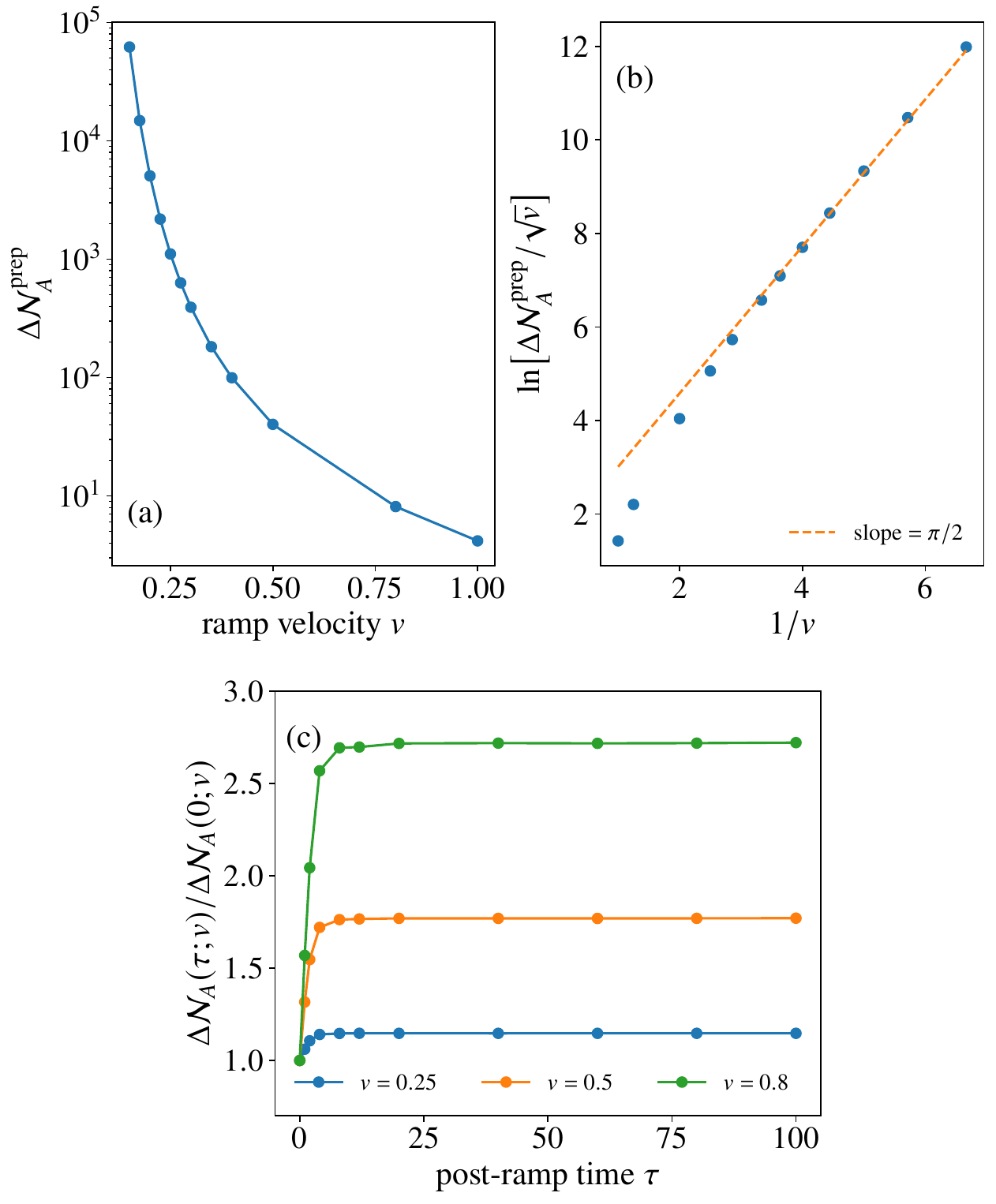}
\caption{
Finite-rate entanglement memory relative to the static real-spectrum endpoint for $\gamma=-0.5$, $\mu_i=-2$, $\mu_f=1.5>\mu_{\rm EP}$, and $\ell=10$.
(a) Prepared-state excess spectral nonpositivity $\Delta\mathcal N_A^{\rm prep} =\mathcal N_A^{\rm prep}-\mathcal N_A^{\rm stat}$ versus ramp velocity. The static endpoint value is $\mathcal N_A^{\rm stat}\simeq0.06593$.
(b) Linearized slow-ramp representation $\ln[\Delta\mathcal N_A^{\rm prep}/\sqrt v]$ versus $1/v$. The dashed line has the fixed predicted slope $\pi\vert{}\gamma\vert{}\Delta^2=\pi/2$, with only its vertical offset fitted.
(c) Post-ramp persistence of the normalized excess $\Delta\mathcal N_A(\tau;v)/\Delta\mathcal N_A(0;v)$ for representative velocities. Over the post-ramp time window shown, the excess remains finite and bounded under the real-spectrum final Hamiltonian rather than relaxing to the static reference.
}
\label{fig:velocity_memory}
\end{figure}

Figure~\ref{fig:velocity_memory}(a) shows $\Delta\mathcal N_A^{\rm prep}$ for $\mu_f=1.5>\mu_{\rm EP}$. The excess grows rapidly in the slow-ramp limit. For these model parameters, $\mathcal A_{\max}=\pi/2\simeq1.571$. In the linearized scaling plot of Fig.~\ref{fig:velocity_memory}(b), lowering the maximum velocity included in the fit from $v=0.3$ to $0.2$ shifts the slope from $\beta_{\rm fit}\simeq1.621$ to $1.590$; intermediate velocity cuts give $1.612$, $1.604$, and $1.597$. This drift moves toward the predicted value $\pi/2\simeq1.571$. The dashed line shows the theoretical slope $\pi/2$ with an adjusted vertical offset, confirming agreement with Eq.~(\ref{eq:velocity_scaling}).

To check whether this excitation survives during post-ramp evolution, we compute $\Delta\mathcal N_A(\tau;v)$ under the real-spectrum Hamiltonian. Figure~\ref{fig:velocity_memory}(c) tracks the ratio $\Delta\mathcal N_A(\tau;v)/\Delta\mathcal N_A(0;v)$ across several ramp velocities. The excess remains pinned near a finite value, displaying weak oscillations without relaxing to zero. We followed this plateau up to hold times $\tau=100$. The finite-rate excess reflects a retained non-Hermitian memory, not the static nonpositivity of the endpoint.

The scaling law does not depend on the specific choice $\gamma=-0.5$ and $\mu_f=1.5$. We performed the same analysis across $\gamma\in\{-0.25,-0.4,-0.5,-0.6,-0.75\}$ using velocity ranges scaled with $\vert{}\gamma\vert{}$, and across endpoints $\mu_f\in\{1.3,1.4,1.5,1.7,2\}$ at fixed $\gamma=-0.5$. Once the sweep clears the broken region, Eq.~(\ref{eq:velocity_scaling}) predicts the slope $\beta_{\mathcal N}=\pi\vert{}\gamma\vert{}\Delta^2$, independent of $\mu_f$. Three-point fits at the slowest velocities yield ratios $\beta_{\mathcal N}/(\pi\vert{}\gamma\vert{}\Delta^2)$ between $0.998$ and $1.013$ across the $\gamma$ series, and between $0.998$ and $1.004$ across the $\mu_f$ series. The scaling holds even though the static baseline shifts substantially with $\mu_f$.

\subsection{Connected-correlation memory}
\label{sec:czz_memory}

Connected longitudinal correlations provide a complementary real-space signature of this memory. Because the state remains Gaussian, $P_k(t)$ fixes all two-point correlators directly. Mapping spin correlations to fermionic contractions relies on standard Jordan--Wigner identities~\cite{LiebSchultzMattis1961,BarouchMcCoy1971}, as applied to driven Ising and XY models~\cite{CherngLevitov2006,Cincio2007,SenguptaSen2009,JafariPRR2025,NajiPRB2025}. Spin-correlation scaling under noisy drives is examined in Ref.~\cite{JafariAkbariCorr2026}. In non-Hermitian Gaussian states, normal and pairing contractions form independent biorthogonal expectation values. Appendix~\ref{app:biorthogonal_correlations} provides the explicit construction.

  In the rotated Nambu frame, the projector components are
\begin{equation}
P_k(t) = \begin{pmatrix} \langle \hat c_k \hat c_k^\dagger\rangle_{LR} & i\,\langle \hat c_k \hat c_{-k}\rangle_{LR} \\ -i\,\langle \hat c_{-k}^\dagger \hat c_k^\dagger\rangle_{LR} & \langle \hat c_{-k}^\dagger \hat c_{-k}\rangle_{LR} \end{pmatrix}.
\label{eq:Pk_correlation_blocks}
\end{equation}
We define the real-space correlators by Fourier transformation:
\begin{equation}
\begin{aligned}
G_r(t)&=\langle \hat c_j^\dagger\hat c_{j+r}\rangle_{LR}, \\ F_r(t)&=\langle \hat c_j\hat c_{j+r}\rangle_{LR}, \qquad 
\overline F_r(t)=\langle \hat c_j^\dagger\hat c_{j+r}^\dagger\rangle_{LR}.
\end{aligned} \label{eq:realspace_contractions}
\end{equation}
Setting $\hat n_j=\hat c_j^\dagger\hat c_j$, the connected density-density correlator is
\begin{equation}
C_{nn}(r,t) \equiv \langle \hat n_j\hat n_{j+r}\rangle_{LR} - \langle \hat n_j\rangle_{LR} \langle \hat n_{j+r}\rangle_{LR}.
\label{eq:Cnn_def}
\end{equation}
For $r\neq0$, Wick factorization yields
\begin{equation}
C_{nn}(r,t) = -\overline F_r(t)F_r(t)-G_r(t)G_{-r}(t),
\label{eq:Cnn_wick}
\end{equation}
and the longitudinal spin correlation evaluates to
\begin{equation}
C^{zz}(r,t)=4 C_{nn}(r,t).
\label{eq:Czz}
\end{equation}
Because $\rho^{LR}$ is not Hermitian, $C^{zz}$ is not constrained by standard unitary bounds. Large values do not reflect normalization errors.

  We apply the same baseline subtraction to the real-space correlator at $\gamma=-0.5$ and $\mu_f=1.5>\mu_{\rm EP}$. We compare the prepared correlation profile against the static reference:
\begin{equation}
\delta C^{zz}(r;v) = C_{\rm prep}^{zz}(r;v) - C_{\rm stat}^{zz}(r).
\label{eq:czz_memory_profile}
\end{equation}
The difference $\delta C^{zz}(r;v)$ extracts the dynamical excitation induced by the ramp. We measure the total magnitude of this memory profile using the $L^2$ norm
\begin{equation}
\begin{aligned}
\mathcal M_C(v) &= \left[ \sum_{r\ge 1} \left| \delta C^{zz}(r;v) \right|^2 \right]^{1/2}.
\end{aligned} \label{eq:czz_memory_measure}
\end{equation}
The numerical sum cuts off at $r_{\max}=80$. Increasing the distance cutoff to $r_{\max}=240$ produces no change within machine precision, showing that the profile is fully contained within the summation window. The following asymptotic formulas evaluate this full spatial norm rather than the fixed-$r_{\max}$ limit.

  The slow-ramp profile is governed by the same accumulated action found in the entanglement sector. Under the present parameters, Eq.~(\ref{eq:ramp_action}) gives
\begin{equation}
\mathcal A(k) = \frac{\pi}{2}\sin^2 k.
\label{eq:czz_action}
\end{equation}
The maxima sit at $k_0=\pm\pi/2$. Expanding around the saddle with $k=k_0+q$ gives
\begin{equation}
\mathcal A(k) = \frac{\pi}{2} - \frac{\pi}{2}q^2 + \mathcal O(q^4).
\label{eq:czz_action_saddle}
\end{equation}
The non-Hermitian amplification in any contraction $X_r\in\{G_r,F_r,\overline F_r\}$ comes from the Fourier integral of $P_k$:
\begin{equation}
\begin{aligned}
X_r^{\rm amp}(v) &\sim \int dk\, f_{X,r}(k,v)\, e^{\mathcal A(k)/v},
\end{aligned} \label{eq:czz_contraction_saddle}
\end{equation}
where $f_{X,r}$ varies algebraically with $v$. At a fixed site separation $r$, standard saddle-point integration yields
\begin{equation}
\begin{aligned}
X_r^{\rm amp}(v) &\sim B_{X,r}\sqrt v\, e^{\mathcal A_{\max}/v}.
\end{aligned} \label{eq:czz_contraction_prefactor}
\end{equation}
Evaluating the full spatial norm requires retaining the coordinate dependence of the saddle. A uniform saddle expansion yields an envelope of the form
\begin{equation}
\begin{aligned}
X_r^{\rm amp}(v) &\sim \sqrt v\, e^{\mathcal A_{\max}/v}\, G_X\!\left( \sqrt v\,[r-r_c(v)] \right),
\end{aligned} \label{eq:czz_contraction_packet}
\end{equation}
indicating a spatial packet width $\Delta r\sim v^{-1/2}$. Because Eq.~(\ref{eq:Cnn_wick}) is bilinear in these contractions, the baseline-subtracted spin correlator scales as
\begin{equation}
\begin{aligned}
\delta C^{zz}(r;v) &\sim v\, e^{2\mathcal A_{\max}/v}\, F\!\left( \sqrt v\,[r-r_c(v)] \right).
\end{aligned} \label{eq:czz_uniform_profile}
\end{equation}
Interference between the two saddle points generates spatial oscillations across the profile. Unless these packets cancel identically, this interference alters the envelope shape without modifying the $v^{-1/2}$ spatial scaling.

Integrating over the spatial profile in Eq.~(\ref{eq:czz_memory_measure}) introduces an extra $v^{-1/2}$ factor from the packet width:
\begin{equation}
\begin{aligned}
\mathcal M_C^2(v) &\sim v^2 e^{4\mathcal A_{\max}/v} \sum_r \left| F\!\left( \sqrt v\,[r-r_c(v)] \right) \right|^2 \\ &\sim v^{3/2}e^{4\mathcal A_{\max}/v} \int dx\,|F(x)|^2 .
\end{aligned} \label{eq:czz_memory_norm_squared}
\end{equation}
Taking the square root establishes the asymptotic scaling for the integrated memory norm:
\begin{equation}
\begin{aligned}
\mathcal M_C(v) &\sim B_0\,v^{3/4} \exp\left[ \frac{2\mathcal A_{\max}}{v} \right].
\end{aligned} \label{eq:czz_memory_prefactor}
\end{equation}
For $\gamma=-0.5$ and $\Delta=1$, this relation takes the form
\begin{equation}
\begin{aligned}
\ln\!\left[ \frac{\mathcal M_C(v)}{v^{3/4}} \right] &= \frac{\pi}{v} +\ln B_0+{o(1)}.
\end{aligned} \label{eq:czz_memory_exponential}
\end{equation}
Evaluating the individual contributions $-\overline F_rF_r$ and $-G_rG_{-r}$ reveals no mutual cancellation. Appendix~\ref{app:slow_ramp_asymptotics} outlines the full-profile derivation. The product of contractions doubles the action in the exponent, while the wavepacket expansion $\Delta r\sim v^{-1/2}$ sets the $v^{3/4}$ prefactor. Compensating the slow-ramp data by $v^{3/4}$ produces a flat plateau, verifying the scaling behavior.

\begin{figure}[t]
\centering
\includegraphics[width=0.48\textwidth]{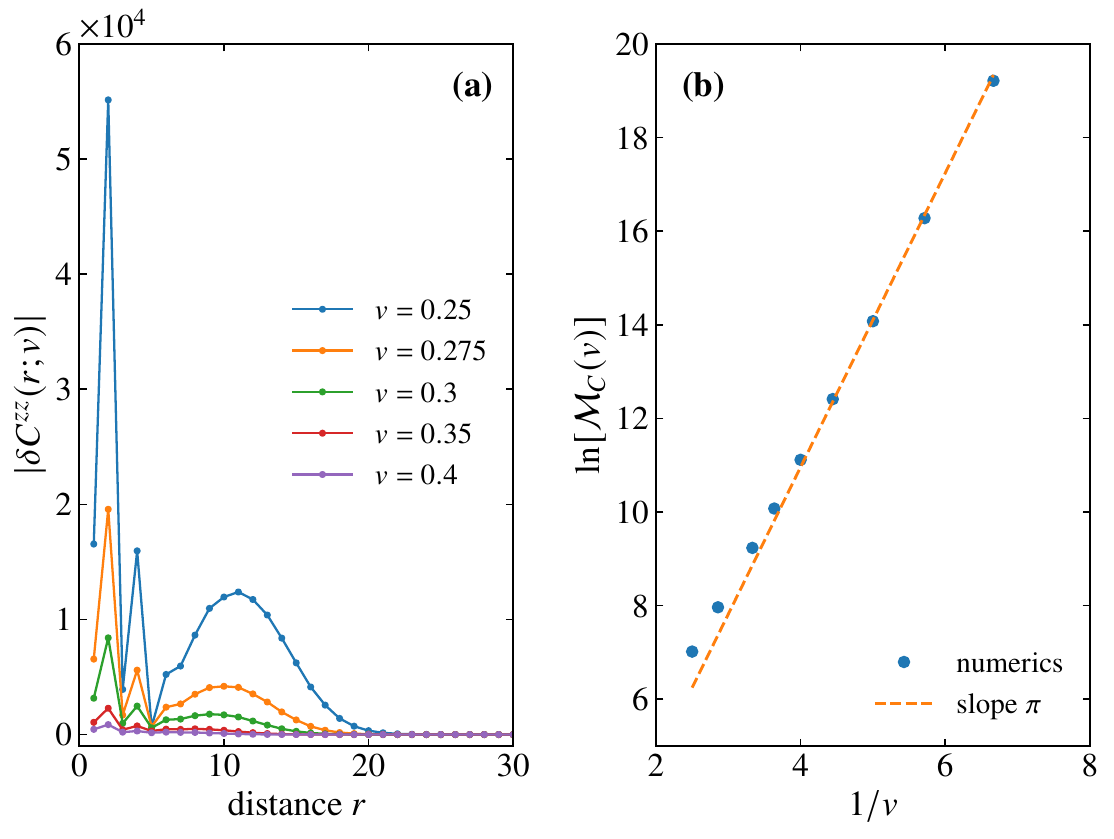}
\caption{Finite-rate longitudinal-correlation memory at the real-spectrum endpoint for $\gamma=-0.5$, {$\mu_i=-2$}, and $\mu_f=1.5>\mu_{\rm EP}$.
(a) Representative real-space memory profiles $\vert{}\delta C^{zz}(r;v)\vert{}$. The integrated memory in Eq.~(\ref{eq:czz_memory_measure}) is evaluated up to $r_{\max}=80$. The oscillatory real-space structure reflects interference between the two symmetry-related amplified saddles at $k_0=\pm\pi/2$, which contribute opposite Fourier phases $e^{\pm i k_0 r}$.
(b) $\ln\mathcal M_C(v)$ versus $1/v$ for the extended high-resolution calculation, including velocities down to $v=0.15$. The dashed line has the fixed slope $\pi$ and displays the leading exponential factor before removal of the algebraic prefactor. The prefactor-corrected linearization $\ln[\mathcal M_C/v^{3/4}]$ is analyzed separately in the text and in Table~\ref{tab:memory_robustness}, using the slow-ramp scan points.}
\label{fig:czz_memory}
\end{figure}

Figure~\ref{fig:czz_memory}(a) illustrates representative correlation profiles. The amplitude increases as the drive velocity is reduced. Figure~\ref{fig:czz_memory}(b) plots the raw logarithm $\ln\mathcal M_C$ against $1/v$ to highlight the doubled exponent directly. Fitting the slope over $v\le0.25$ gives $\beta\simeq3.043$; restricting the range to $v\le0.20$ yields $\beta\simeq3.083$, converging toward the predicted exponent $2\mathcal A_{\max}=\pi\simeq3.142$.

The power-law prefactor is removed in the slow-ramp fit. For $\gamma=-0.5$ and $\mu_f=1.5$, fitting $\ln[\mathcal M_C/v^{3/4}]$ across the slowest points ($v=0.10, 0.12, 0.14$) yields $\beta_C\simeq3.1458$, matching $\pi$ within $0.13\%$. Table~\ref{tab:memory_robustness} compiles the three-point fits: the ratio $\beta_C/(2\pi\vert{}\gamma\vert{}\Delta^2)$ spans $1.001$--$1.014$ across the imbalance scan and $1.001$--$1.010$ across the endpoint scan. Accounting for the packet width confirms the doubled action while leaving the underlying amplification intact. This profile represents a non-equilibrium excess created by the drive history.

\begin{table}[b]
\caption{Robustness of the slow-ramp action slopes using the three slowest velocities in each fit. We define $R_{\mathcal N}\equiv\beta_{\mathcal N}/(\pi\vert{}\gamma\vert{}\Delta^2)$ from $\ln[\Delta\mathcal N_A^{\rm prep}/\sqrt v]$ and $R_C\equiv\beta_C/(2\pi\vert{}\gamma\vert{}\Delta^2)$ from $\ln[\mathcal M_C/v^{3/4}]$. For the endpoint scan, $\gamma=-0.5$ and all $\mu_f$ lie in the real-spectrum regime.}
\label{tab:memory_robustness}
\begin{ruledtabular}
\begin{tabular}{lccc}
scan &
range &
$R_{\mathcal N}$ &
$R_C$
\\
$\gamma$ &
$[-0.75,-0.25]$ &
$0.998$--$1.013$ &
$1.001$--$1.014$
\\
$\mu_f$ &
$[1.30,2.00]$ &
$0.998$--$1.004$ &
$1.001$--$1.010$
\end{tabular}
\end{ruledtabular}
\end{table}

\subsection{Contrast: drive ending in the broken-spectrum region}

  Ramps ending within the broken-spectrum region trigger ongoing post-ramp amplification. For an unstable mode in the final Hamiltonian, we denote the imaginary dispersion by $\varepsilon_k(\mu_f)=i\kappa_k$, where $\kappa_k=\vert{}\operatorname{Im}\varepsilon_k(\mu_f)\vert{}>0$. In Eq.~(\ref{eq:postramp_projector}), the interband components carry the time dependence $e^{-i(E_{+,k}-E_{-,k})\tau}=e^{2\kappa_k\tau}$. The momentum-resolved growth rate is
\begin{equation}
\Gamma_k \equiv 2\kappa_k =2|\operatorname{Im}\varepsilon_k(\mu_f)|.
\label{eq:kitaev_growth_rate}
\end{equation}
Near a quadratic maximum,
\begin{equation}
\Gamma_k = \Gamma_{\max} -a(k-k_0)^2+\cdots, \qquad a>0,
\label{eq:kitaev_growth_saddle}
\end{equation}
and integrating over momentum produces the long-time asymptotic form $\tau^{-1/2}e^{\Gamma_{\max}\tau}$. This runaway growth contrasts with the saturated memory found at real-spectrum endpoints.

  Consider $\gamma=-0.5$ and $\mu_f=1$. The dispersion satisfies
\begin{equation}
\varepsilon_k^2 = (\cos k-1)^2-\frac12\sin^2k = \frac32\cos^2k-2\cos k+\frac12 .
\label{eq:eps2_muf1}
\end{equation}
The imaginary component peaks at $\cos k=2/3$, giving $\vert{}\operatorname{Im}\varepsilon_k\vert{}=1/\sqrt6$. From Eq.~(\ref{eq:kitaev_growth_rate}), the maximal growth rate evaluates to
\begin{equation}
\Gamma_{\max} = \Gamma_{\rm th} = \sqrt{\frac{2}{3}}.
\label{eq:Gamma_th}
\end{equation}
The two momenta $k_0=\pm\arccos(2/3)$ realize this supremum. Expanding around either point ($k=k_0+q$) yields $\Gamma_k=\Gamma_{\rm th}[1-(5/2)q^2+\mathcal O(q^3)]$, confirming a parabolic saddle.

\begin{figure}[t]
\centering
\includegraphics[width=0.48\textwidth]{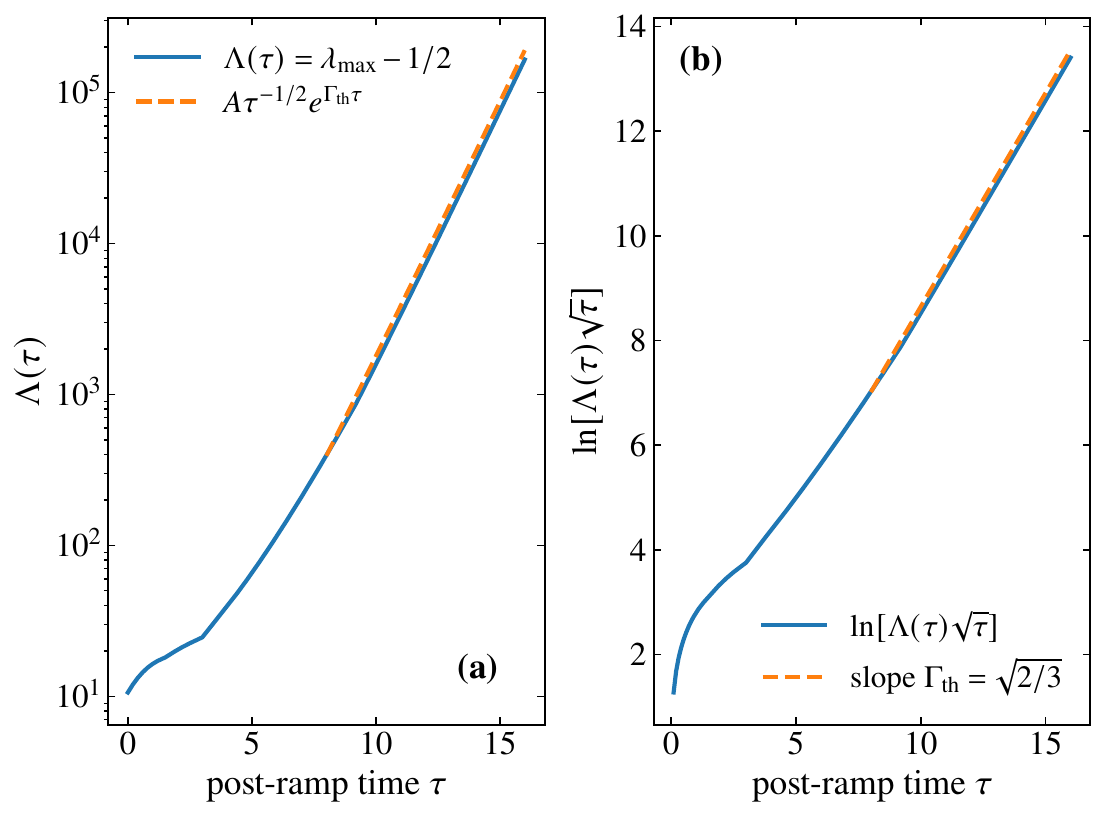}
\caption{
Broken-spectrum amplification for $\gamma=-0.5$, $\mu_i=-2$, $\mu_f=1$, $v=0.5$, and $\ell=10$.
(a) Long-time growth of $\Lambda(\tau)$, defined in Eq.~(\ref{eq:Lambda_def}), together with the predicted asymptotic behavior $\tau^{-1/2}e^{\sqrt{2/3}\tau}$.
(b) Linearized representation $\ln[\Lambda(\tau)\sqrt{\tau}]$, whose late-time slope approaches $\sqrt{2/3}$. The dashed line has the fixed theoretical slope $\Gamma_{\rm th}=\sqrt{2/3}$, with only its vertical offset fitted.
}
\label{fig:broken_growth}
\end{figure}

  The momentum integral gives the one-dimensional factor $\tau^{-1/2}e^{\Gamma_{\rm th}\tau}$. By Eq.~(\ref{eq:CA_spectral_constraint}), the subsystem entanglement spectrum remains real even though the post-quench Hamiltonian is complex. Pairing symmetry from Eq.~(\ref{eq:subsystem_fermionic_compatibility}) enforces $\lambda_{\max}=1-\lambda_{\min}$. We define the spectral spread
\begin{equation}
\begin{aligned}
\Lambda(\tau) \equiv \frac{
\lambda_{\max}(\tau)
\!-\!
\lambda_{\min}(\tau)}{2} = \lambda_{\max}(\tau)
\!-\!
\frac12 = \frac12
\!-\!
\lambda_{\min}(\tau).
\end{aligned} \label{eq:Lambda_def}
\end{equation}
measuring the distance of the extremal eigenvalues from $1/2$.
  When the momentum saddle dominates the projection onto subsystem $A$ without destructive interference, $\Lambda(\tau)$ follows the bulk saddle scaling:
\begin{equation}
\begin{aligned}
\Lambda(\tau) \propto \tau^{-1/2} e^{\Gamma_{\rm th}\tau} = \tau^{-1/2} e^{\sqrt{2/3}\,\tau}.
\end{aligned} \label{eq:Lambda_growth}
\end{equation}
Figure~\ref{fig:broken_growth} confirms this behavior. Fitting $\ln[\Lambda(\tau)\sqrt{\tau}]$ over the window $10\le\tau\le16$ yields $\Gamma_{\rm fit}\simeq0.8146$, close to $\Gamma_{\rm th}=\sqrt{2/3}\simeq0.8165$. Shifting the lower fitting limit from $\tau=8$ to $10$ changes the extracted rate from $0.8061$ to $0.8146$, providing an estimate of preasymptotic finite-time corrections.

\section{Exceptional-endpoint response}
\label{sec:exceptional}

\subsection{Finite-time regularity and the exact halt}
\label{sec:finite_endpoint}

  Entanglement and correlation properties near exceptional points have been documented in non-Hermitian spin models~\cite{Miao2024}. Here we consider driven protocols and approach the positive exceptional boundary from the gapped real-spectrum side, setting $\mu_f=\mu_{\rm EP}+\delta\mu$ with $\delta\mu>0$. Below, $S_A$ designates the continuously tracked entropy $S_A^{\rm cont}$ unless stated otherwise. At the exceptional momentum $k_{\rm EP}$, we define the positive quasiparticle energy of the target Hamiltonian by $\varepsilon_{\rm EP}(\delta\mu)\equiv \varepsilon_{k_{\rm EP}}(\mu_f=\mu_{\rm EP}+\delta\mu)$. Equation~(\ref{eq:kep}) yields
\begin{equation}
\varepsilon_{\rm EP}^{\,2}(\delta\mu) = \delta\mu^2 - \frac{2\gamma\Delta^2}{\mu_{\rm EP}}\, \delta\mu .
\label{eq:eps_ep_exact}
\end{equation}
Taking $\gamma<0$ and expanding as $\delta\mu\to0^+$ gives the standard square-root closure,
\begin{equation}
\varepsilon_{\rm EP}(\delta\mu) = \sqrt{-\frac{2\gamma\Delta^2}{\mu_{\rm EP}}}\, \delta\mu^{1/2} + \mathcal O(\delta\mu^{3/2}).
\label{eq:eps_ep_sqrt}
\end{equation}

Finite-time regularity does not rely on specific two-band algebraic identities. 
For any finite-dimensional Hamiltonian that depends analytically on a control parameter, the matrix exponential $e^{-iH\tau}$ is analytic in that parameter at fixed finite $\tau$. The matrix exponential is an entire function of its matrix argument, so defective Hamiltonians and Puiseux-type eigenvalue branchings do not disrupt this operator analyticity.

For the two-band problem, the identity $H_k^2=\varepsilon_k^2 I$ makes this regularity elementary. The post-ramp propagator takes the closed form
\begin{equation}
\begin{aligned}
U_k(\tau) &= \cos(\varepsilon_k\tau)I -i\frac{\sin(\varepsilon_k\tau)}{\varepsilon_k}H_k .
\end{aligned} \label{eq:exact_propagator}
\end{equation}
Both $\cos(\varepsilon_k\tau)$ and $\sin(\varepsilon_k\tau)/\varepsilon_k$ are entire functions of $\varepsilon_k^2$. Because both $\varepsilon_k^2$ and $H_k$ vary analytically with $\delta\mu$, the evolution operator $U_k(\tau)$ is analytic in $\delta\mu$ across all momenta, including the exceptional mode. At $k=k_{\rm EP}$, the target Hamiltonian reads $H_f=H_{\rm EP}-\delta\mu\,\sigma_z$. Expanding Eq.~(\ref{eq:eps_ep_exact}) for finite $\tau$ gives
\begin{equation}
\begin{aligned}
U_{k_{\rm EP}}(\delta\mu,\tau) &= I-iH_{\rm EP}\tau +\mathcal O(\delta\mu).
\end{aligned} \label{eq:U_regular}
\end{equation}
The quasiparticle gap closes as $\sqrt{\delta\mu}$. The finite-time propagator, by contrast, contains no such branch point.

  Right at the exceptional boundary, the generator is nilpotent with $H_{\rm EP}^2=0$. The propagator terminates at the first Jordan order:
\begin{equation}
U_{\rm EP}(\tau) = e^{-iH_{\rm EP}\tau} = I-iH_{\rm EP}\tau .
\label{eq:Jordan_U}
\end{equation}
Equation~(\ref{eq:U_regular}) converges continuously to this linear Jordan evolution for any bounded post-ramp duration $\tau$.

The finite-rate ramp exhibits the same smoothness. Each choice of $\delta\mu$ shifts the upper ramp boundary to $\mu_f=\mu_{\rm EP}+\delta\mu$, adding a duration $\delta\mu/v$ to the protocol. For fixed ramp rate $v>0$, the time-ordered evolution operator depends analytically on the integration upper limit. The matrix elements of $H_k(\mu)$ stay analytic across $k$ and $\mu$. Because the initial band gap guarantees a regular projector at $\mu_i$, and the Brillouin zone and ramp path are compact, this analyticity is uniform in momentum. Fourier transformation and subsystem restriction then give
\begin{equation}
\begin{aligned}
\mathcal C_A(\delta\mu;v,\tau) &= \mathcal C_A^{\rm halt}(v,\tau) + \delta\mu\,\mathcal X_A(v,\tau) + \mathcal O(\delta\mu^2), \\ \mathcal X_A(v,\tau) &\equiv \left. \frac{\partial\mathcal C_A(\delta\mu;v,\tau)} {\partial(\delta\mu)} \right|_{\delta\mu=0^+}.
\end{aligned} \label{eq:CA_linear_endpoint}
\end{equation}

   We evaluate the distance from the defective limit using the entrywise Frobenius metric:
\begin{equation}
\begin{aligned}
D_C(\delta\mu,\tau) &\equiv \left\| \mathcal C_A(\delta\mu;v,\tau) - \mathcal C_A^{\rm halt}(v,\tau) \right\|_F, \\ \|Y\|_F &\equiv \sqrt{\operatorname{Tr}\!\left(Y^\dagger Y\right)} = \left[ \sum_{ij}|Y_{ij}|^2 \right]^{1/2}.
\end{aligned} \label{eq:frobenius_distance}
\end{equation}
The metric $D_C$ measures matrix deviations directly. It avoids Riemann surface ambiguities and vanishes only when the correlation matrices match identically.

Individual eigenvalues may lose analytic parameterizations near defective degeneracies even when matrix elements remain regular. The expansion for the entropy is therefore conditional. Provided an analytic branch exists and no eigenvalue crosses $0$ or $1$, the subsystem entropy satisfies
\begin{equation}
\begin{aligned}
S_A(\delta\mu;v,\tau) &= S_A^{\rm halt}(v,\tau) + b_A(v,\tau)\,\delta\mu + \mathcal O(\delta\mu^2), \\ b_A(v,\tau) &\equiv \left. \frac{\partial S_A(\delta\mu;v,\tau)} {\partial(\delta\mu)} \right|_{\delta\mu=0^+}.
\end{aligned} \label{eq:S_linear_endpoint}
\end{equation}

\begin{figure}[t]
\centering
\includegraphics[width=0.48\textwidth]{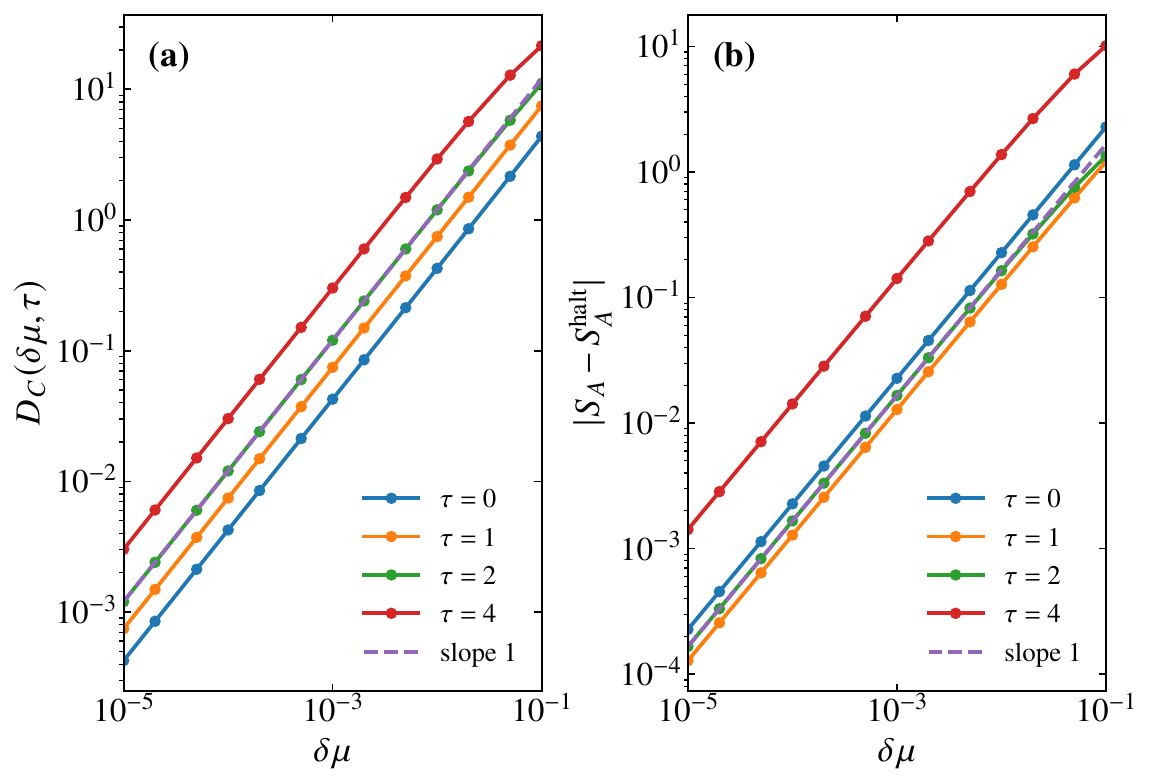}
\caption{
Finite-time approach to the exceptional endpoint for $\gamma=-0.5$, $\mu_i=-2$, $v=0.8$, and $\ell=10$, with $\mu_f=\mu_{\rm EP}+\delta\mu$.
(a) Frobenius distance $D_C(\delta\mu,\tau) =\Vert\mathcal C_A(\delta\mu;v,\tau) -\mathcal C_A^{\rm halt}(v,\tau)\Vert_F$ from the exact-halt correlation matrix.
(b) Corresponding continuously followed entropy difference $\vert{}S_A(\delta\mu;v,\tau)-S_A^{\rm halt}(v,\tau)\vert{}$. For every fixed post-ramp time shown, both quantities approach the exact-halt limit linearly in $\delta\mu$. The dashed lines indicate linear scaling.
}
\label{fig:endpoint_scaling}
\end{figure}

Figure~\ref{fig:endpoint_scaling} tests this expansion for $v=0.8$. The correlation matrix converges to the exact-halt limit with an explicit linear $\delta\mu$ scaling. The tracked entropy reproduces this linear slope on the chosen sheet. Fits over $10^{-4}\leq\delta\mu\leq10^{-2}$ yield power-law exponents indistinguishable from unity across all tested hold times $\tau$. The square-root branch cut in the Hamiltonian spectrum produces no matching branch cut in the finite-time correlation dynamics.

For finite rings with antiperiodic boundary conditions, $k_{\rm EP}$ rarely coincides with an allowed crystal momentum. The nilpotent Jordan block is strictly a property of the thermodynamic Bloch continuum. Individual isolated momenta carry measure zero in this limit; physical deviations in subsystem observables stem from momentum intervals surrounding $k_{\rm EP}$ rather than from isolated points.

\subsection{Long-time exceptional crossover}
\label{sec:longtime}

The linear dependence established above holds only at fixed finite $\tau$. As the drive terminates closer to the boundary, the band splitting narrows as $\varepsilon_{\rm EP}\sim\sqrt{\delta\mu}$. The dynamics is governed by the scaling combination
\begin{equation}
\begin{aligned}
x &= \varepsilon_{\rm EP}(\delta\mu)\tau \sim \tau\sqrt{\delta\mu}.
\end{aligned} \label{eq:double_scaling_x}
\end{equation}
For $x\ll1$, the linear Taylor expansion in $\delta\mu$ holds. Once $x\sim 1$, the time evolution resolves the small quasiparticle gap. The characteristic crossover timescale scales as the inverse splitting:
\begin{equation}
\begin{aligned}
\tau_{\rm EP} &\sim \varepsilon_{\rm EP}^{-1} \sim \delta\mu^{-1/2}.
\end{aligned} \label{eq:tau_EP}
\end{equation}
  We analyze this crossover using the matrix distance $D_C(\delta\mu,\tau)$ from Eq.~(\ref{eq:frobenius_distance}). We construct the normalized deviation
\begin{equation}
Q_C(\delta\mu,\tau) = \frac{D_C(\delta\mu,\tau)/\delta\mu} {\displaystyle \lim_{\delta\mu'\to0^+} D_C(\delta\mu',\tau)/\delta\mu'}.
\label{eq:QC}
\end{equation}
In the early-time linear regime, $Q_C\to1$. We evaluate the denominator numerically from small-$\delta\mu$ slopes at each hold time $\tau$.

\begin{figure}[t]
\centering
\includegraphics[width=\columnwidth]{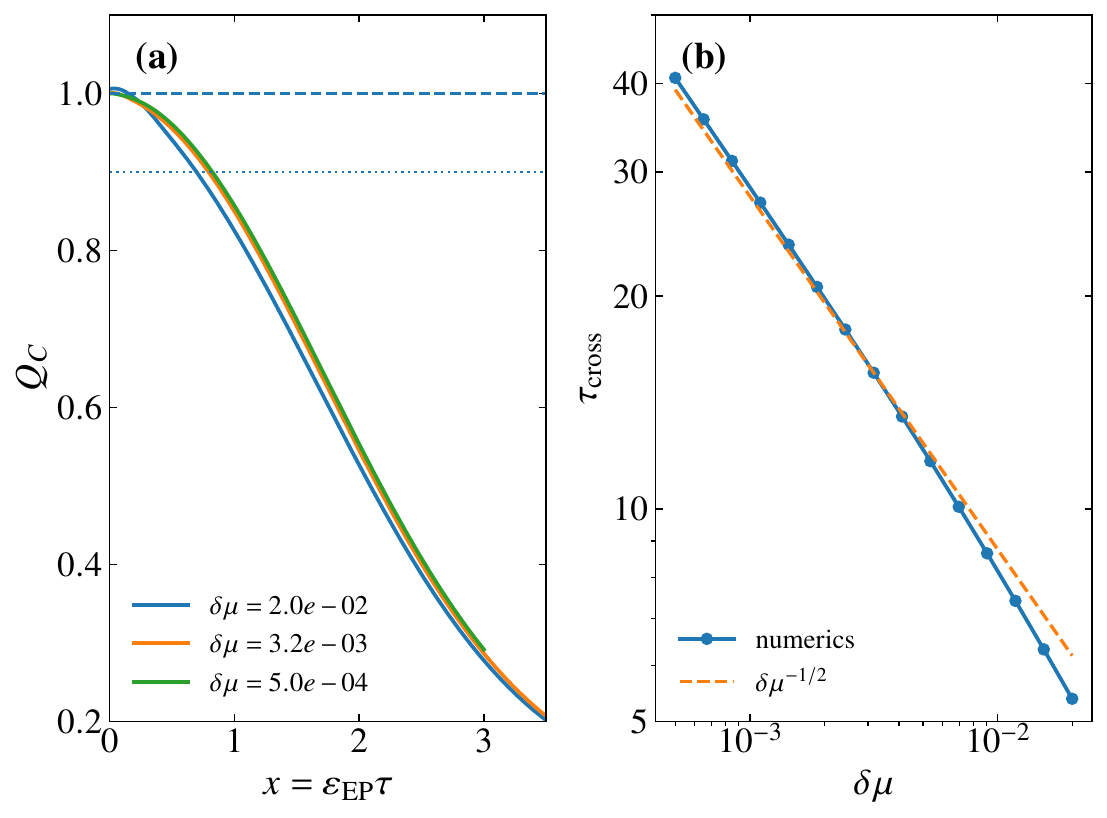}
\caption{
Long-time crossover near the exceptional endpoint for $\gamma=-0.5$, $\mu_i=-2$, $v=0.8$, and $\ell=10$.
(a) Normalized correlation-matrix deviation $Q_C$ plotted against $x=\varepsilon_{\rm EP}\tau$ for several endpoint distances $\delta\mu$. The near collapse shows that the crossover is controlled by $x$. The dashed line marks the fixed-time limit $Q_C=1$, while the dotted line marks the $10\%$ deviation criterion used to define $\tau_{\rm cross}$.
(b) Extracted crossover time $\tau_{\rm cross}$ versus $\delta\mu$. The dashed line shows the predicted scaling $\tau_{\rm cross}\propto\delta\mu^{-1/2}$.
}
\label{fig:longtime_EP_crossover}
\end{figure}

\begin{figure}[t]
\centering
\includegraphics[width=\columnwidth]{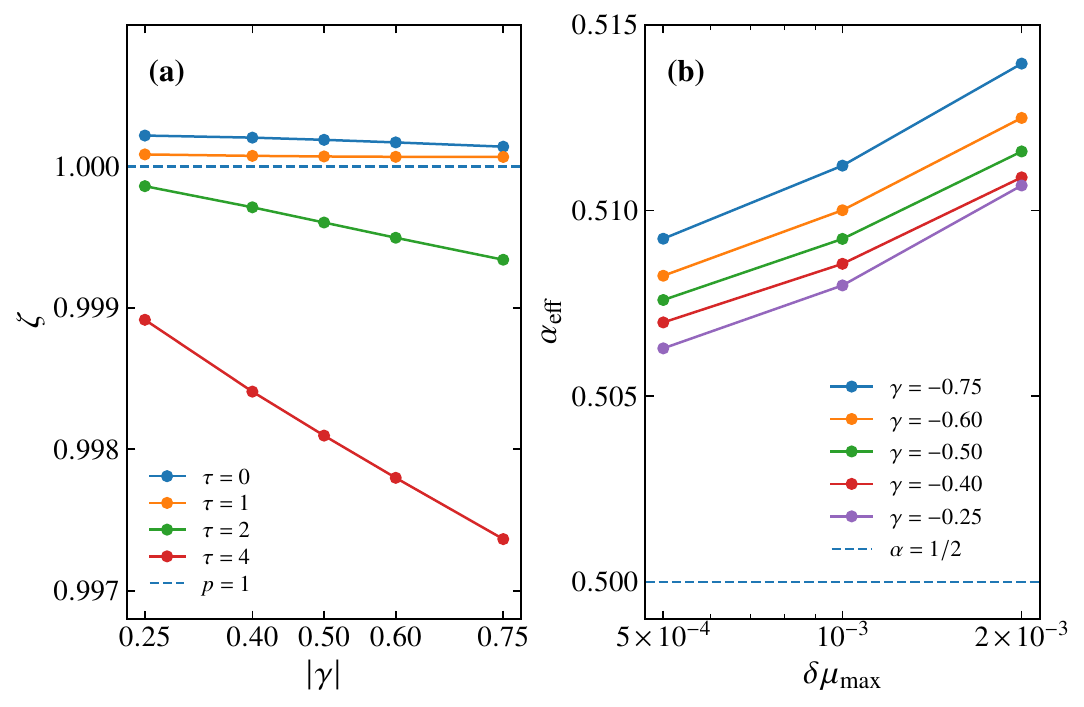}
\caption{Robustness of the exceptional-endpoint scaling for {$\mu_i=-2$}.
(a) Fitted finite-time exponent $\zeta$ in $D_C\propto(\delta\mu)^\zeta$ versus $\vert{}\gamma\vert{}$ for several fixed post-ramp times. The dashed line marks the predicted linear value $\zeta=1$.
(b) Effective crossover exponent $\alpha_{\rm eff}$ from $\tau_{\rm cross}\propto(\delta\mu)^{-\alpha_{\rm eff}}$ versus the largest endpoint distance $\delta\mu_{\max}$ included in the fit. As the fitting window is restricted toward the exceptional endpoint, the curves approach the asymptotic exponent $\alpha=1/2$, shown by the dashed line.}
\label{fig:EP_robustness}
\end{figure}

Figure~\ref{fig:longtime_EP_crossover} demonstrates data collapse for distinct values of $\delta\mu$ plotted against $x=\varepsilon_{\rm EP}\tau$. Because this crossover is continuous, we identify the crossover point $\tau_{\rm cross}$ through the condition $\vert{}Q_C(\delta\mu,\tau_{\rm cross})-1\vert{}=0.1$, setting a $10\%$ departure from early-time linearity. Choosing a different numerical percentage shifts the prefactor of $\tau_{\rm cross}$ without altering the $\delta\mu^{-1/2}$ exponent.
For the tightest endpoint distance, $\delta\mu=5\times10^{-4}$, we checked stability up to $x\le3.5$ ($\tau\le173$). Correlation distances computed with momentum discretizations $N_k=4096$ and $8192$ agree within relative errors below $10^{-12}$ across this entire range. Momentum resolution is not an issue here. The square-root energy scale is absent from early-time derivatives, but governs the divergence of the long-time scale $\tau_{\rm EP}$.
These scaling exponents hold away from $\gamma=-0.5$. For the early-time response, fits of $D_C\propto(\delta\mu)^\zeta$ across $\gamma\in\{-0.25,-0.4,-0.5,-0.6,-0.75\}$ and $\tau\in\{0,1,2,4\}$ with $\delta\mu\le2\times10^{-3}$ yield exponents $0.9974\lesssim \zeta\lesssim1.0002$. As illustrated in Fig.~\ref{fig:EP_robustness}(a), the exponent $\zeta$ stays locked to unity across the negative-imbalance phase.
For the long-time crossover, applying the $10\%$ threshold across the same parameter scan $\gamma\in\{-0.25,-0.4,-0.5,-0.6,-0.75\}$ produces effective exponents $\alpha_{\rm eff}\simeq0.511$--$0.514$ over the broad parameter window. Fitting only the three smallest values of $\delta\mu$ shifts the exponent to $\alpha_{\rm eff}\simeq0.506$--$0.509$. Figure~\ref{fig:EP_robustness}(b) confirms convergence toward the asymptotic value $\alpha=1/2$, showing that deviations reflect preasymptotic corrections.

\section{Spatial dynamics at the exceptional halt}
\label{sec:spatial_dynamics}

\subsection{Correlation fronts}
\label{sec:spatial_fronts}

\begin{figure}[t]
\centering
\includegraphics[width=0.48\textwidth]{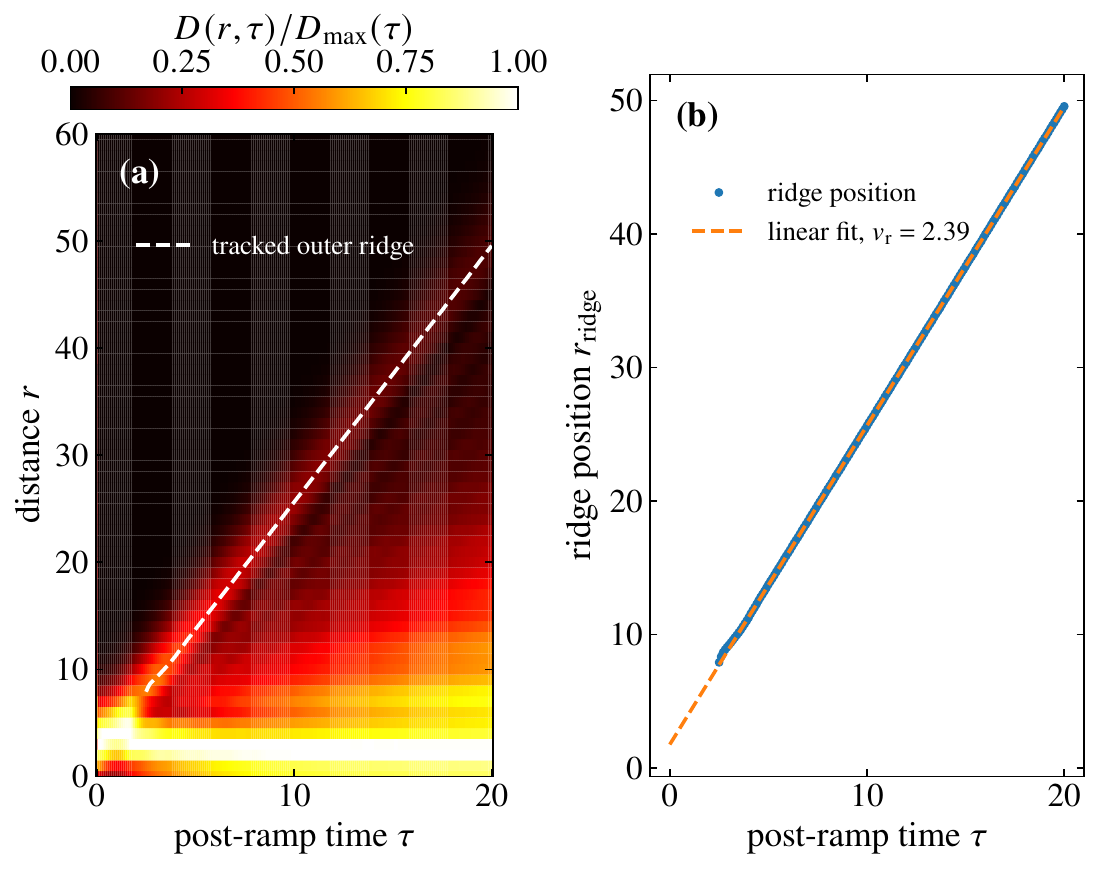}
\caption{
Real-space correlation-projector spreading at the exact exceptional halt for $\gamma=-0.5$, {$\mu_i=-2$}, and $v=0.8$.
(a) Normalized correlation-matrix change $D(r,\tau)/D_{\max}(\tau)$, with the tracked outer ridge shown by the dashed line.
(b) Ridge position as a function of post-ramp time. The dashed line is a linear fit, giving $v_{\rm ridge}\simeq2.39$. The fit remains stable when the early-time part is excluded.
}
\label{fig:realspace_ridge}
\end{figure}

We track the real-space correlation-projector front directly. In standard Hermitian one-dimensional setups, post-quench propagation is ballistic and governed by free quasiparticle velocities~\cite{CalabreseCardy2005}. By contrast, exceptional non-Hermitian quenches can produce supersonic wavefronts alongside multiple nested light cones~\cite{BacsiDora2021}. The propagation behavior at an exact exceptional halt cannot be assumed a priori; it must be calculated.
  Fourier transforming the momentum-space correlation projector yields
\begin{equation}
\mathcal P_r(\tau)
=
\frac{1}{N_k}
\sum_k e^{ikr}P_k(\tau),
\end{equation}
The post-ramp spatial deviation from the halt configuration is measured by the Frobenius distance
\begin{equation}
D(r,\tau) = \left\| \mathcal P_r(\tau)-\mathcal P_r(0) \right\|_F . \label{eq:realspace_distance}
\end{equation}
For profile normalization, let $D_{\max}(\tau)\equiv\max_r D(r,\tau)$.
The space-time profile of $D(r,\tau)$ in Fig.~\ref{fig:realspace_ridge} displays a well-defined outer ridge moving linearly outward. Tracking this local maximum across hold times yields
\begin{equation}
r_{\rm ridge}(\tau) \simeq r_{0,\rm ridge}+v_{\rm ridge}\tau, \qquad v_{\rm ridge}\simeq2.39. \label{eq:ridge_velocity}
\end{equation}
This linear fit is robust. Truncating the early transient interval ($\tau < 2$) alters the fitted slope by less than one percent.

  At the exceptional boundary and for the parameters chosen here, the dispersion reduces to
\begin{equation}
\varepsilon_k^2 = \frac{3}{2} \left( \cos k-\sqrt{\frac{2}{3}} \right)^2 .
\end{equation}
Off-diagonal projector matrix elements rotate with the interband phase factor generated by $2\varepsilon_k$. Stationary-phase arguments therefore fix the maximal group velocity through
\begin{equation}
v_{\max} = \max_k \left| \frac{\partial(2\varepsilon_k)}{\partial k} \right| = 2\max_k \left| \frac{\partial\varepsilon_k}{\partial k} \right| = \sqrt{6} \simeq2.45 . \label{eq:max_correlation_velocity}
\end{equation}
The maximum in Eq.~(\ref{eq:max_correlation_velocity}) is realized at $k=\pi/2$, well away from the exceptional mode $k_{\rm EP}$. Thus, $v_{\max}$ reflects the global band dispersion rather than local exceptional-point kinematics, without invoking a Lieb-Robinson bound.
The extracted velocity $v_{\rm ridge}\simeq 2.39$ closely approaches this upper limit. The projector spreads ballistically. Unlike the multiple light cones observed in quenched non-Hermitian SSH chains~\cite{BacsiDora2021}, only a single dominant wavefront emerges here.
This distinction involves several technical factors. Ref.~\cite{BacsiDora2021} constructs its correlation matrix from a single right state normalized under the standard Dirac norm. We use the overlap-normalized biorthogonal density matrix with an inverse-propagated left dual. Because the driving protocol, lattice Hamiltonian, and inner-product conventions all differ, the absence of multiple fronts cannot be attributed to any single difference in isolation.

\begin{figure}[t]
\centering
\includegraphics[width=0.48\textwidth]{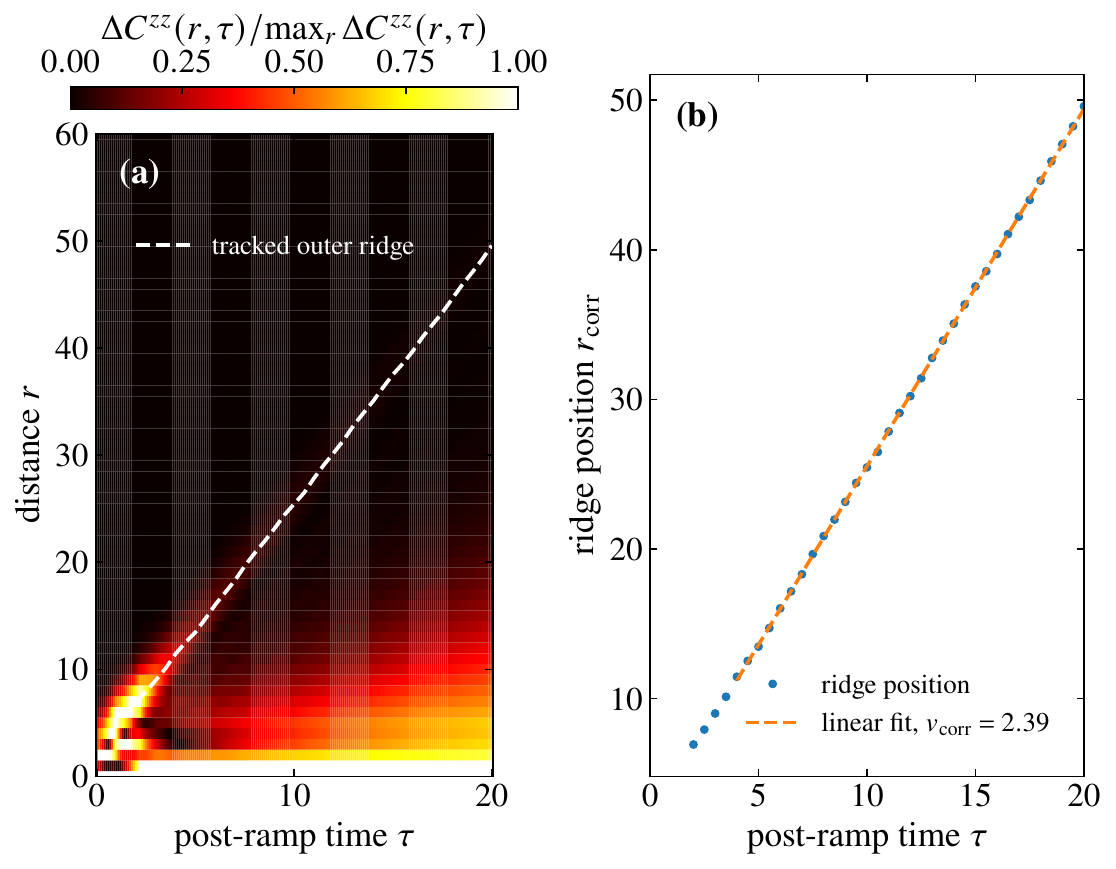}
\caption{
Biorthogonal connected longitudinal-correlation spreading at the exact exceptional halt for $\gamma=-0.5$, {$\mu_i=-2$}, and $v=0.8$.
(a) Normalized post-ramp change $\Delta C^{zz}(r,\tau)/\max_r\Delta C^{zz}(r,\tau)$, with the tracked outer ridge shown by the dashed line.
(b) Ridge position as a function of post-ramp time. The dashed line is a linear fit over $\tau\ge4$, giving $v_{\rm corr}\simeq2.39$.
}
\label{fig:czz_front}
\end{figure}

  The connected longitudinal spin correlator provides an independent check. At the same exceptional endpoint, we evaluate the post-ramp deviation profile
\begin{equation}
\Delta C^{zz}(r,\tau) = \left| C^{zz}(r,\tau)-C^{zz}(r,0) \right|. \label{eq:DeltaCzz}
\end{equation}
This observable is independent of logarithm branch cuts and maps directly onto the single-particle projector front.
Figure~\ref{fig:czz_front} shows the space-time propagation of $\Delta C^{zz}(r,\tau)$ for $\gamma=-0.5$ and $v=0.8$. Locating the outer edge across time yields
\begin{equation}
r_{\rm corr}(\tau) \simeq r_{0,\rm corr}+v_{\rm corr}\tau, \qquad v_{\rm corr}\simeq2.39 . \label{eq:czz_velocity}
\end{equation}
Restricting the fit to $\tau\ge4$ yields $v_{\rm corr}\simeq2.386$, remaining stable against variations in momentum grid size and ramp step resolution. This matches the projector velocity $v_{\rm ridge}\simeq2.39$ and lies close to the band bound $v_{\max}=\sqrt6\simeq2.45$. The two-point observable confirms the ballistic scale determined from the projector.

\subsection{Subsystem saturation}
\label{sec:subsystem_scaling}

We examine the subsystem scaling at the exact exceptional boundary $\mu_f=\mu_{\rm EP}$. Because the biorthogonal entropy becomes complex, we evaluate its principal-branch real component, $\operatorname{Re}S_A^{\rm pr}$, alongside the spectral nonpositivity metric $\mathcal N_A$. The reality of the underlying single-particle spectrum at the halt is guaranteed by the pseudo-Hermitian relation in Eq.~(\ref{eq:CA_spectral_constraint}).

  At fixed post-ramp times, both quantities saturate as the block length $\ell$ increases, defining an expanding spatial boundary. We extract this saturation scale $\ell_{\rm sat}$ by parameterizing the curves with a continuous piecewise linear-to-plateau envelope:
\begin{equation}
Q_A(\ell,\tau) = \begin{cases} Q_\infty(\tau) + m(\tau)\,[\ell-\ell_{\rm sat}(\tau)], & \ell<\ell_{\rm sat}(\tau), \\[1mm] Q_\infty(\tau), & \ell\ge\ell_{\rm sat}(\tau), \end{cases}
\label{eq:saturation_fit}
\end{equation}
where $Q_A$ represents either $\operatorname{Re}S_A^{\rm pr}$ or $\mathcal N_A$. The plateau level is $Q_\infty(\tau)$, and $m(\tau)$ is the slope of the sub-saturation branch.
Equation~(\ref{eq:saturation_fit}) provides a convenient operational metric for the crossover point; it does not represent an exact non-analytic transition. It fits the numerical saturation smoothly for all hold times, improving as $\tau$ grows.

\begin{figure}[t]
\centering
\includegraphics[width=0.48\textwidth]{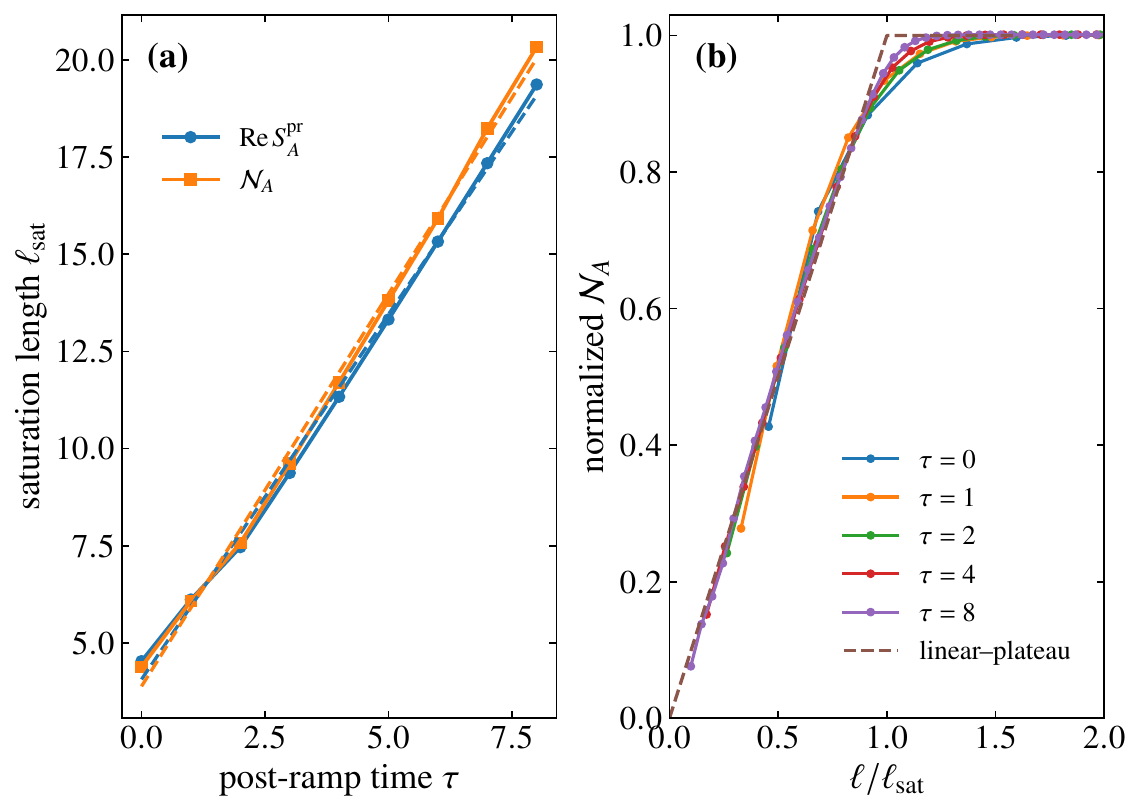}
\caption{
Subsystem-size saturation at the exact exceptional halt for $\gamma=-0.5$, {$\mu_i=-2$}, and $v=0.8$.
(a) Saturation length $\ell_{\rm sat}$ extracted independently from $\operatorname{Re}S_A^{\rm pr}$ and $\mathcal N_A$. The dashed lines are linear fits, $\ell_{\rm sat}=\ell_0+u\tau$, with $u\simeq1.88$ and $2.02$, respectively; here $u$ is an operational saturation-length growth rate, not a front velocity.
(b) Normalized spectral nonpositivity $\mathcal N_A(\ell,\tau)/\mathcal N_A^\infty(\tau)$ after rescaling the subsystem size by $\ell_{\rm sat}$, where $\mathcal N_A^\infty(\tau)$ is the fitted plateau value. The curves show an approximate collapse through the crossover region; the dashed line is the linear-to-plateau form used as a guide to the eye.
}
\label{fig:subsystem_saturation}
\end{figure}

  Figure~\ref{fig:subsystem_saturation} plots the resulting length scales. The values extracted from $\operatorname{Re}S_A^{\rm pr}$ and $\mathcal N_A$ track one another closely and expand linearly:
\begin{equation}
\ell_{\rm sat}(\tau) = \ell_0+u\tau .
\label{eq:saturation_length_growth}
\end{equation}
Fitting across nine uniform points ($\tau=0,1,\ldots,8$) gives $u\simeq1.88$ for $\operatorname{Re}S_A^{\rm pr}$ and $u\simeq2.02$ for $\mathcal N_A$. Dropping the initial point ($\tau\ge1$) gives $u\simeq1.93$ and $2.07$.
To verify that this scale is not an artifact of the piecewise model, we evaluate a model-independent fractional length $\ell_f$, defined as the length where the observable completes a fraction $f$ of its total climb from $\ell=1$ to the large-$\ell$ asymptote. Taking thresholds $f=0.80$, $0.90$, and $0.95$ recovers linear growth. For $\tau\ge1$, the corresponding slopes lie in the intervals $1.50$--$1.81$ for $\operatorname{Re}S_A^{\rm pr}$ and $1.62$--$1.94$ for $\mathcal N_A$. The expansion of the subsystem boundary is genuine, though the effective growth velocity depends on the extraction rule and should not be confused with a microscopic quasiparticle speed.

Rescaling the subsystem size by $\ell_{\rm sat}$ collapses the nonpositivity profiles onto a universal crossover curve. In Fig.~\ref{fig:subsystem_saturation}(b), we plot $\mathcal N_A(\ell,\tau)/\mathcal N_A^\infty(\tau)$, where $\mathcal N_A^\infty(\tau)$ is the plateau amplitude. A control calculation deep in the real-spectrum phase ($\mu_f=1.5>\mu_{\rm EP}$) exhibits the same behavior: the $90\%$ metric for $\tau\ge1$ yields slopes of roughly $1.62$ for $\operatorname{Re}S_A^{\rm pr}$ and $1.75$ for $\mathcal N_A$. Spatial saturation is generic to the finite-rate preparation; it is not a unique signature of the exceptional point.
The spatial position of the correlation ridge matches the saturation length closely once the profile is established. At $\tau=4$, we find $r_{\rm ridge}\simeq11.31$, compared to $\ell_{\rm sat}\simeq11.33$ from the entropy and $11.70$ from nonpositivity. At $\tau=8$, the corresponding lengths are $20.87$, $19.36$, and $20.33$. This geometric agreement does not imply identical asymptotic velocities. The linear intercepts differ, and the slope $u$ remains consistently below the correlation velocity $v_{\rm ridge}\simeq2.39$. The subsystem saturation length and the correlation wavefront represent distinct, non-identical spatial scales expanding in parallel.
\section{Summary and outlook}
\label{sec:summary}

Our results separate non-Hermitian amplification during broken-spectrum passages from the spectral singularities of exceptional endpoints in finite-time driven dynamics. Transient sweeps across complex-energy bands leave residual excitations that survive long after the Hamiltonian returns to an unbroken, real-spectrum domain. Because static ground states at negative imbalance already host spectral nonpositivity, this static baseline must be subtracted out to isolate the genuine non-equilibrium contribution. Real-space connected longitudinal correlations exhibit the identical drive-history dependence, confirming that this memory is not an artifact of the entanglement formulation. Both the entanglement excess and the correlation norm follow the integrated imaginary action across all tested values of pairing imbalance and real target endpoints; the full-profile correlation norm acquires a spatial wavepacket factor $v^{3/4}$ alongside the doubled action exponent. Halting the sweep inside the broken-spectrum interval instead leaves unstable sectors to amplify indefinitely.

Exceptional endpoints operate under a different mechanism.
The square-root vanishing of the instantaneous quasiparticle gap does not induce a singular finite-time response in reduced-state observables. The finite-duration propagator and the subsystem correlation matrix converge to the defective endpoint with linear regularity. Spectral singularities emerge only in the asymptotic long-time regime via a crossover from early-time operator analyticity to late-time gap sensitivity. Both the linear approach exponent and the crossover scaling remain invariant throughout the negative-imbalance regime.
At the exact exceptional boundary, the single-particle correlation projector and the connected longitudinal spin correlation spread behind an identical ballistic wavefront. The subsystem saturation length sets a comparable but quantitatively distinct expanding spatial scale. Positive pairing imbalance maps to a Hermitian Kitaev chain via a local similarity transformation, supplying an exact benchmark for the driven biorthogonal dynamics. Moving beyond Gaussian systems to explore how transient amplification and exceptional degeneracies interact in the presence of many-body interactions remains an open problem.

\section*{Acknowledgments}

A.~A. acknowledges financial support from the Beijing Natural Science
Foundation under Grant No.~IS25015.
The work of H.~Y. was supported by the Beijing Natural Science Foundation
under Grant No.~IS23013.


\appendix

\section{Gaussian entropy and logarithm branches}
\label{app:entropy_branches}

  Diagonalizable Gaussian reduced operators decompose into independent single-particle sectors. In biorthogonal fermionic normal form, the paired single-particle spectrum partitions into $\ell$ distinct modes~\cite{PeschelEisler2009,Herviou2019,Guo2021}. Choosing one representative eigenvalue $\lambda_\alpha$ from each complementary set $\{\lambda_\alpha, 1-\lambda_\alpha\}$, the many-body eigenvalues of $\rho_A^{LR}$ factorize into a product:
\begin{equation}
w_{\{n_\alpha\}}
=
\prod_{\alpha=1}^{\ell}
\lambda_\alpha^{\,n_\alpha}
(1-\lambda_\alpha)^{1-n_\alpha},
\qquad
n_\alpha=0,1.
\label{eq:manybody_gaussian_weights}
\end{equation}
The logarithm required for the Gaussian entropy evaluates additively mode by mode:
\begin{equation}
\operatorname{Log}_{G} w_{\{n_\alpha\}}
=
\sum_{\alpha=1}^{\ell}
\left[
n_\alpha\operatorname{Log}\lambda_\alpha
+
(1-n_\alpha)\operatorname{Log}(1-\lambda_\alpha)
\right].
\label{eq:gaussian_log}
\end{equation}
We term Eq.~(\ref{eq:gaussian_log}) the additive Gaussian logarithm prescription.

For Hermitian states, this definition matches the conventional von Neumann entropy. Without Hermiticity, it departs from a single principal matrix logarithm of the full operator $\rho_A^{LR}$ whenever eigenvalues become complex or negative. A real eigenvalue $\lambda_\alpha\notin[0,1]$ yields negative many-body spectral weights; this generates the spectral nonpositivity analyzed in the main text.

The static variant $S_A^{\rm pr}$ evaluates each logarithm in Eq.~(\ref{eq:general_entropy_spectrum}) on the principal sheet with branch cut $(-\pi,\pi]$. For negative real arguments, this sets $\operatorname{Log}(-x)=\ln x+i\pi$ with $x>0$. Because each time step is treated independently, a mode crossing the negative real axis produces a step discontinuity. The dynamical entropy $S_A^{\rm cont}$ resolves this artifact. It initializes on the principal branch at $t=0$, tracking each mode continuously across Riemann sheets as the drive proceeds.

We match eigenvalues between adjacent time slices by minimizing pairwise Euclidean distances, unwrapping the accumulated phase angles of $\lambda_\nu$ and $1-\lambda_\nu$ along smooth parametric trajectories. Machine-precision imaginary components do not dictate branch crossings. When an eigenvalue hits the points $0$ or $1$, we enforce the regular limit $z\operatorname{Log}z\to0$ and inherit the branch index directly from the preceding time step. For strictly real spectra, Eq.~(\ref{eq:ImS_nonpositivity}) provides an exact consistency check on the chosen branches.


\section{Metric constraint and reality of the subsystem spectrum}
\label{app:spectral_reality}

  We prove the spectral constraints stated in Sec.~\ref{sec:memory_baseline} for the positive-energy preparation and inverse-propagated dual frame. For real $\gamma<0$, the pseudo-Hermiticity relation in Eq.~(\ref{eq:eta_pseudo_hermiticity}) gives
\begin{equation}
U_k^\dagger(t)\eta U_k(t)=\eta .
\label{eq:eta_pseudounitary}
\end{equation}
Deep in the initial real-spectrum sector ($\mu_i<-\mu_{\rm EP}$), one has $\mu_k>0$ and $0<\varepsilon_k<\mu_k$. The positive-energy spectral projector is
\begin{equation}
P_k^+
=
\frac{1}{2}
\left(
I+\frac{H_k}{\varepsilon_k}
\right).
\label{eq:positive_projector_metric}
\end{equation}
The matrix $\eta P_k^+$ is Hermitian and has rank one. Its semidefinite sign is fixed by its trace:
\begin{equation}
\operatorname{Tr}(\eta P_k^+)
=
\frac{1}{2}
\left[
1+\gamma
+
(\gamma-1)\frac{\mu_k}{\varepsilon_k}
\right]
<\gamma<0 .
\label{eq:left_metric_trace}
\end{equation}
The complementary projector $I-P_k^+$ carries the opposite sign.
This sets the initial bounds:
\begin{equation}
\eta P_k(0)\preceq0,
\qquad
\eta[I-P_k(0)]\succeq0 .
\label{eq:initial_metric_signs}
\end{equation}
Combining $P_k(t)=U_k(t)P_k(0)U_k^{-1}(t)$ with the pseudo-unitary relation in Eq.~(\ref{eq:eta_pseudounitary}) yields
\begin{equation}
\eta P_k(t)
=
U_k^{-\dagger}(t)\,
\eta P_k(0)\,
U_k^{-1}(t),
\label{eq:metric_congruence_P}
\end{equation}
with an identical congruence holding for $I-P_k(t)$. Congruence preserves matrix inertia. The semidefinite inequalities therefore hold for all times $t$, through and beyond the complex-energy window.

The real-space representation is reached by direct sum: $\bigoplus_k P_k$. The Fourier transform is unitary and commutes with the site-independent metric $I\otimes\eta$. Real-space operators inherit these semidefinite bounds; restricting to subsystem $A$ preserves them as principal submatrices. Defining
\begin{equation}
\eta_A=
\begin{pmatrix}
\gamma I_\ell&0\\
0&I_\ell
\end{pmatrix},
\end{equation}
one obtains the block inequalities
\begin{equation}
\eta_A\mathcal C_A\preceq0,
\qquad
\eta_A(I-\mathcal C_A)\succeq0 .
\label{eq:subsystem_metric_inequalities}
\end{equation}
Reality follows by similarity. Let $Q=-\eta_A\mathcal C_A\succeq0$. The nonzero spectrum of $\mathcal C_A=-\eta_A^{-1}Q$ matches that of the Hermitian matrix
\begin{equation}
-Q^{1/2}\eta_A^{-1}Q^{1/2},
\end{equation}
forcing the eigenvalues of $\mathcal C_A$ to remain strictly real.
  Eigenvalues within the open interval $(0,1)$ are forbidden. Let $\mathcal C_Ax=\lambda x$ with $0<\lambda<1$. Contracting $x$ against Eq.~(\ref{eq:subsystem_metric_inequalities}) requires both $x^\dagger\eta_Ax\le0$ and $x^\dagger\eta_Ax\ge0$, so $x^\dagger\eta_Ax=0$. This forces the quadratic form to vanish:
\begin{equation}
x^\dagger Qx
=
-\lambda x^\dagger\eta_Ax
=
0 .
\end{equation}
Because $Q\succeq0$, it follows that $Qx=0$, implying $\mathcal C_Ax=-\eta_A^{-1}Qx=0$, which contradicts $\lambda>0$. This completes the proof of Eq.~(\ref{eq:CA_spectral_constraint}).

On the right side ($\mu_f>\mu_{\rm EP}$), the inequalities in Eq.~(\ref{eq:initial_metric_signs}) reverse sign for the static positive-energy projector; an identical argument establishes Eq.~(\ref{eq:CA_spectral_constraint}) for the static reference state. The proof relies on initializing on the positive-energy branch and propagating the left state with the inverse dual. It does not generalize to arbitrary initial states.

\section{Slow-ramp saddle asymptotics}
\label{app:slow_ramp_asymptotics}

  For complete crossings, the single-particle action in Eq.~(\ref{eq:ramp_action}) takes the form
\begin{equation}
\begin{aligned}
\mathcal A(k) &= \mathcal A_{\max}\sin^2 k, \qquad \mathcal A_{\max} = \pi|\gamma|\Delta^2 .
\end{aligned} \label{eq:app_action_compact}
\end{equation}
The integrated momentum envelope evaluates analytically to
\begin{equation}
\begin{aligned}
\mathcal J(v) \equiv \int_{-\pi}^{\pi} \frac{dk}{2\pi} \exp\!\left[ \frac{\mathcal A(k)}{v} \right] = \exp\!\left[ \frac{\mathcal A_{\max}}{2v} \right] I_0\!\left( \frac{\mathcal A_{\max}}{2v} \right),
\end{aligned} \label{eq:app_Bessel_exact}
\end{equation}
where $I_0$ denotes the modified Bessel function of the first kind. Expanding for small velocity $v\to0$ gives
\begin{equation}
\begin{aligned}
\mathcal J(v) &= \frac{\sqrt v} {\pi|\Delta|\sqrt{|\gamma|}} \exp\!\left[ \frac{\mathcal A_{\max}}{v} \right] \left[ 1+ \frac{v}{4\mathcal A_{\max}} + \mathcal O(v^2) \right]. \end{aligned}
\label{eq:app_Bessel_asymptotic}
\end{equation}
The $\sqrt v$ factor originates from the twin Gaussian saddles at $k_0=\pm\pi/2$. This factor controls any observable with a nonvanishing saddle coefficient.

  For a finite subsystem, assume the nonexponential matrix connection factor is regular and retains a nonzero restricted projection. The amplified part of the correlation matrix scales as
\begin{equation}
\begin{aligned}
\delta\mathcal C_A(v) &= \mathcal J(v) \left[ \mathcal K_A^{(0)} + \mathcal O(v) \right],
\end{aligned} \label{eq:app_CA_uniform_saddle}
\end{equation}
where $\mathcal K_A^{(0)}$ encodes connection coefficients and spatial boundaries. Let $\kappa_\nu$ denote its nonzero eigenvalues, and define
\begin{equation}
\begin{aligned}
\mathcal Q_A &\equiv \frac{1}{2} \sum_\nu |\kappa_\nu|.
\end{aligned} \label{eq:app_QA}
\end{equation}
By the spectral reality established in Appendix~\ref{app:spectral_reality}, the exponential growth of $\mathcal J(v)$ renders the static baseline negligible as $v\to0$. If $\mathcal Q_A\neq0$, the prepared excess scales as
\begin{equation}
\begin{aligned}
\Delta\mathcal N_A^{\rm prep}(v) &\sim A_0\sqrt v\, e^{\mathcal A_{\max}/v}, \qquad A_0 = \frac{\mathcal Q_A} {\pi|\Delta|\sqrt{|\gamma|}} .
\end{aligned} \label{eq:app_A0_explicit}
\end{equation}
The plateau observed in the compensated numerical data confirms $\mathcal Q_A\neq0$ across the parameters examined in the main text.

The saddle width sets the spatial scaling of the two-point functions. Retaining the stationary-phase coordinate dependence gives the profile
\begin{equation}
\begin{aligned}
\delta X_r(v) &\sim \sqrt v\, e^{\mathcal A_{\max}/v} \, \mathcal G_X\!\left( \sqrt v\,[r-r_c(v)] \right), \end{aligned} \label{eq:app_uniform_contraction}
\end{equation}
with $X_r\in\{G_r,F_r,\overline F_r\}$. Quadratic Wick contractions therefore scale in amplitude as $v e^{2\mathcal A_{\max}/v}$, spread over a spatial packet $\Delta r\sim v^{-1/2}$. Integrating this profile into the full spatial norm produces
\begin{equation}
\begin{aligned}
\mathcal M_C(v) &\sim B_0\,v^{3/4} e^{2\mathcal A_{\max}/v}.
\end{aligned} \label{eq:app_B0_explicit}
\end{equation}
The nonuniversal constant $B_0$ captures connection weights, spatial phases, saddle interference, and Wick factors. Compensated slow-ramp data verify that $B_0$ is finite and nonzero across the tested range. The exponential action and power-law exponent stem from saddle geometry; $A_0$ and $B_0$ carry the remaining nonuniversal details.

\section{Operator form of the correlation projector for the Kitaev chain}
\label{app:kitaev_projector_operator}

  For the rotated Nambu basis of Sec.~\ref{sec:kitaev_model}, $\widetilde\Psi_k = \left( \hat c_k,\, -i\hat c_{-k}^{\dagger} \right)^T$, the operator dyadic in Eq.~(\ref{eq:Pk_manybody_relation}) reads
\begin{equation}
\widetilde\Psi_k\widetilde\Psi_k^\dagger
=
\begin{pmatrix}
\hat c_k\hat c_k^\dagger
&
i\,\hat c_k\hat c_{-k}
\\[1mm]
-i\,\hat c_{-k}^\dagger\hat c_k^\dagger
&
\hat c_{-k}^\dagger\hat c_{-k}
\end{pmatrix}.
\label{eq:kitaev_operator_outer_product}
\end{equation}
Taking the normalized biorthogonal expectation value generates
\begin{equation}
P_k(t)
=
\begin{pmatrix}
\langle \hat c_k\hat c_k^\dagger\rangle_{LR}
&
i\,\langle \hat c_k\hat c_{-k}\rangle_{LR}
\\[1mm]
-i\,\langle \hat c_{-k}^\dagger\hat c_k^\dagger\rangle_{LR}
&
\langle \hat c_{-k}^\dagger\hat c_{-k}\rangle_{LR}
\end{pmatrix},
\label{eq:kitaev_operator_expectation}
\end{equation}
which is the momentum-space correlation projector evaluated in the text.

  At $t=0$, let $\vert{}R_{+,k}(0)\rangle$ and $\langle L_{+,k}(0)\vert{}$ denote the right and left positive-energy eigenvectors of $H_k(0)$, normalized such that $\langle L_{+,k}(0)\vert{}R_{+,k}(0)\rangle=1$. The initial Gaussian state gives
\begin{equation}
P_k(0)
=
|R_{+,k}(0)\rangle
\langle L_{+,k}(0)|.
\label{eq:kitaev_initial_spectral_projector}
\end{equation}
During the drive, the projector updates via
\begin{equation}
P_k(t)
=
U_k(t)P_k(0)U_k^{-1}(t),
\label{eq:kitaev_driven_projector_appendix}
\end{equation}
in agreement with Eq.~(\ref{eq:general_similarity_evolution}). The matrix $P_k(t)$ remains the correlation projector of the propagated state; it does not match the instantaneous spectral projector of $H_k(t)$. Propagated vectors do not track instantaneous eigenvectors. The factor $U_k^{-1}$ replaces $U_k^\dagger$ because the left dual evolves under the inverse generator.

\section{Biorthogonal real-space correlation functions}
\label{app:biorthogonal_correlations}

We detail the real-space contractions and connected longitudinal correlators used in Sec.~\ref{sec:czz_memory}. This construction adapts the Jordan--Wigner and Wick techniques established for integrable spin systems~\cite{LiebSchultzMattis1961,BarouchMcCoy1971} and driven Ising/XY lattices~\cite{CherngLevitov2006,Cincio2007,SenguptaSen2009,JafariPRR2025,NajiPRB2025}. Spin correlations in non-Hermitian XY models are discussed in Ref.~\cite{Miao2024}. The modification here is that expectation values follow the normalized biorthogonal pairing of Eq.~(\ref{eq:LR_expectation}); the anomalous pairing functions are not complex conjugates.

  Converting momentum contractions to real space proceeds via standard discrete Fourier transforms. In a translationally invariant Gaussian state, the momentum contractions are
\begin{equation}
\begin{aligned}
n_k \!\equiv\! \langle \hat c_k^\dagger\hat c_k\rangle_{LR}, \qquad f_k \!\equiv\! \langle \hat c_k\hat c_{-k}\rangle_{LR}, \qquad \overline f_k \!\equiv\! \langle \hat c_k^\dagger\hat c_{-k}^\dagger\rangle_{LR}.
\end{aligned} \label{eq:app_momentum_contractions}
\end{equation}
Using the basis of Eq.~(\ref{eq:Pk_correlation_blocks}), these quantities are extracted from projector elements:
\begin{equation}
\begin{aligned}
n_k = 1-\left[P_k\right]_{11}, \qquad f_k = -i\left[P_k\right]_{12}, \qquad \overline f_k = -i\left[P_k\right]_{21}.
\end{aligned} \label{eq:app_projector_components}
\end{equation}
The final equality uses the identity $\langle \hat c_{-k}^\dagger\hat c_k^\dagger\rangle_{LR}
=-\langle \hat c_k^\dagger\hat c_{-k}^\dagger\rangle_{LR}$. The lower diagonal element satisfies $[P_k]_{22}=n_{-k}$.

  Fourier transformation gives four closed fermionic two-point correlators:
\begin{equation}
\begin{aligned}
\langle \hat c_j^\dagger\hat c_{j+r}^\dagger \rangle_{LR} &= \overline F_r = -\frac{i}{N} \sum_k e^{ikr} \left[P_k\right]_{21}, \\ \langle \hat c_j\hat c_{j+r} \rangle_{LR} &= F_r = -\frac{i}{N} \sum_k e^{-ikr} \left[P_k\right]_{12}, \\ \langle \hat c_j^\dagger\hat c_{j+r} \rangle_{LR} &= G_r = \frac{1}{N} \sum_k e^{ikr} \left( 1-\left[P_k\right]_{11} \right), \\ \langle \hat c_j\hat c_{j+r}^\dagger \rangle_{LR} &= \frac{1}{N} \sum_k e^{-ikr} \left[P_k\right]_{11} = \delta_{r0}-G_{-r}.
\end{aligned} \label{eq:app_closed_fermion_correlators}
\end{equation}
Matrix elements of $P_k$ are evaluated in the basis of Eq.~(\ref{eq:Pk_correlation_blocks}). These relations provide the biorthogonal analogs of the closed forms used in Refs.~\cite{JafariPRR2025,NajiPRB2025}. Building real-space correlators requires summing over the complete Brillouin zone from $(k,-k)$ pairs. Anticommutation implies $F_{-r}=-F_r$ and $\overline F_{-r}=-\overline F_r$. Contractions are algebraically independent; standard Hermitian constraints such as $\overline F_r=-F_r^*$ or $G_{-r}=G_r^*$ do not apply.

  To connect with standard Jordan--Wigner formulations, define the Majorana-like operators
\begin{equation}
\begin{aligned}
A_j &= \hat c_j^\dagger+\hat c_j, \qquad B_j = \hat c_j^\dagger-\hat c_j .
\end{aligned} \label{eq:app_AB_definition}
\end{equation}
Their biorthogonal contractions read
\begin{equation}
\begin{aligned}
\langle A_jA_{j+r}\rangle_{LR} &= \overline F_r+F_r+G_r-G_{-r}+\delta_{r0}, \\ \langle B_jB_{j+r}\rangle_{LR} &= \overline F_r+F_r-G_r+G_{-r}-\delta_{r0}, \\ \langle A_jB_{j+r}\rangle_{LR} &= \overline F_r-F_r-G_r-G_{-r}+\delta_{r0}, \\ \langle B_jA_{j+r}\rangle_{LR} &= \overline F_r-F_r+G_r+G_{-r}-\delta_{r0}. \end{aligned}
\label{eq:app_AB_correlations}
\end{equation}
All equal-time spin observables reduce to linear combinations of $G_r$, $F_r$, and $\overline F_r$. Equations~(\ref{eq:app_closed_fermion_correlators}) and~(\ref{eq:app_AB_correlations}) play the role of the standard $A$--$B$ algebra from Refs.~\cite{JafariPRR2025,NajiPRB2025}. In driven biorthogonal dynamics, these functions cannot be simplified to a single scalar excitation probability; off-diagonal elements remain independent dynamical fields.

Because the left and right states are Gaussian vacua with nonvanishing overlap, their contractions obey Wick's theorem~\cite{BalianBrezin1969}. Parity symmetry eliminates single-fermion expectations. An even product $\phi_1\phi_2\cdots\phi_{2m}$ of operators chosen from the set $\{A_j, B_j\}$ evaluates to a Pfaffian:
\begin{equation}
\begin{aligned}
\left\langle \phi_1\phi_2\cdots\phi_{2m} \right\rangle_{LR}
\!=\!
\operatorname{Pf}\Gamma,
\qquad
{\Gamma_{ab}
\!=\!
\begin{cases}
\langle\phi_a\phi_b\rangle_{LR}
& a<b,
\\
0
& a=b
,\\
-\langle\phi_b\phi_a\rangle_{LR}
& a>b.
\end{cases}}
\end{aligned}
\label{eq:app_biorthogonal_pfaffian}
\end{equation}
This establishes the biorthogonal counterpart of the Pfaffian determinant methods used in XY chains~\cite{LiebSchultzMattis1961,BarouchMcCoy1971,CherngLevitov2006,Cincio2007,JafariPRR2025,NajiPRB2025}.

   For longitudinal spin correlators, the expression simplifies. Taking $\hat n_j=\hat c_j^\dagger\hat c_j$ with $r\neq0$, Wick factorization gives
\begin{equation}
\begin{aligned}
\langle \hat n_j\hat n_{j+r} \rangle_{LR} &= \langle \hat n_j\rangle_{LR} \langle \hat n_{j+r}\rangle_{LR} 
- \langle \hat c_j^\dagger\hat c_{j+r}^\dagger \rangle_{LR} \langle \hat c_j\hat c_{j+r} \rangle_{LR} \\ &\quad + \langle \hat c_j^\dagger\hat c_{j+r} \rangle_{LR} \langle \hat c_j\hat c_{j+r}^\dagger \rangle_{LR} \\ &= G_0^2 - \overline F_rF_r - G_rG_{-r}.
\end{aligned} \label{eq:app_density_wick_derivation}
\end{equation}
Subtracting the product of local densities yields
\begin{equation}
\begin{aligned}
C_{nn}(r,t) &= -\overline F_r(t)F_r(t) -G_r(t)G_{-r}(t), \quad r\neq0,
\end{aligned} \label{eq:app_Cnn_closed}
\end{equation}
which recovers Eq.~(\ref{eq:Cnn_wick}). At zero separation, $\hat n_j^2=\hat n_j$ gives $C_{nn}(0)=G_0-G_0^2$.

   The longitudinal spin operator is $\sigma_j^z=s(2\hat n_j-1)$ with $s=\pm1$. Because $s^2=1$, the connected correlation does not depend on the sign convention:
\begin{equation}
\begin{aligned}
C^{zz}(r,t) &\equiv \langle \sigma_j^z\sigma_{j+r}^z\rangle_{LR} - \langle \sigma_j^z\rangle_{LR} \langle \sigma_{j+r}^z\rangle_{LR} \\ &= \left\langle (2\hat n_j-1)(2\hat n_{j+r}-1) \right\rangle_{LR} \\ &\quad - \left( 2\langle\hat n_j\rangle_{LR}-1 \right) \left( 2\langle\hat n_{j+r}\rangle_{LR}-1 \right) \\ &= 4\left[ \langle\hat n_j\hat n_{j+r}\rangle_{LR} - \langle\hat n_j\rangle_{LR} \langle\hat n_{j+r}\rangle_{LR} \right] \\ &= 4\,C_{nn}(r,t) \\ &= -4 \left[ \overline F_r(t)F_r(t) + G_r(t)G_{-r}(t) \right], \quad r\neq0 .
\end{aligned} \label{eq:app_Czz_closed}
\end{equation}
Equation~(\ref{eq:app_Czz_closed}) provides the analytical form used in the numerical calculations. It makes explicit that the longitudinal observable is quadratic in single-particle amplitudes, which doubles the exponential action exponent discussed in Sec.~\ref{sec:czz_memory}.

\section{Numerical details and convergence}
\label{app:numerics}

During the ramp, we propagate the right and left Nambu vectors independently
using a fourth-order Runge--Kutta method with $\mu$ as the integration
variable. Unless stated otherwise, we use a maximum step size
$|\Delta\mu|=10^{-3}$. Every 25 steps, and again at the end of the ramp,
we restore $\langle L_k|R_k\rangle=1$ and rescale the two vectors
reciprocally. This leaves the projector $P_k$ unchanged. After the ramp,
the Hamiltonian is time independent, and the evolution is evaluated
directly with the exact propagator in Eq.~(\ref{eq:exact_propagator}),
so no additional time discretization is needed.

We monitor the conditioning of each momentum sector through
\begin{equation}
\mathcal K_k=
\frac{\|R_k\|\,\|L_k\|}
{|\langle L_k|R_k\rangle|}.
\label{eq:condition_measure}
\end{equation}
We also monitor the condition number of the eigenvector matrix of
$\mathcal C_A$ during the entropy calculations.

The main results were checked by varying both the momentum grid and the
ramp step. For Fig.~\ref{fig:broken_growth}, increasing the momentum
resolution from $N_k=1024$ to $4096$ and reducing the ramp step by a
factor of four produced no visible change in the results. The prepared-state
excess in Fig.~\ref{fig:velocity_memory} was similarly stable when the
momentum grid was increased up to $N_k=8192$ and the ramp step was reduced.
The correlation-front calculations were checked by changing both the
momentum resolution and the ramp step, while the slow-ramp scans were
verified by doubling $N_k$ and halving the step size. For the long-time
exceptional crossover, the smallest endpoint distance was additionally
checked using $N_k=8192$. These tests confirm that the numerical results
reported in the main text are stable under the numerical refinements used
here.

\bibliography{References}

\end{document}